\documentclass[letter]{IEEEtran}

\usepackage[utf8]{inputenc} 
\usepackage[T1]{fontenc}
\usepackage{url}
\usepackage{ifthen}
\usepackage{cite}
\usepackage[cmex10]{amsmath} % Use the [cmex10] option to ensure complicance
\usepackage{amssymb} % including amsfonts
\usepackage{dsfont}

\ifCLASSINFOpdf
  \usepackage[pdftex]{graphicx}
\else
  \usepackage[dvips]{graphicx}
\fi
\usepackage{kotex}
\usepackage{xcolor}
\usepackage{subfigure}
\usepackage{tikz}
\usetikzlibrary{calc, arrows.meta, positioning, fit, decorations.pathreplacing}

\usepackage[ruled, linesnumbered]{algorithm2e}
\usepackage{etoolbox}
\makeatletter
\patchcmd{\@algocf@start}% <cmd>
  {-1.5em}% <search>
  {0pt}% <replace>
  {}{}% <success><failure>
\makeatother
\usepackage{lipsum}

\newtheorem{remark}{Remark}

\usepackage{stackengine}
\DeclareMathOperator*{\defeq}{\mathrel{\ensurestackMath{\stackon[1pt]{=}{\scriptscriptstyle\Delta}}}}

\newcommand\cA{{\mathcal{A}}}  \newcommand\cC{{\mathcal{C}}}   \newcommand\cF{{\mathcal{F}}}   \newcommand\cI{{\mathcal{I}}}     \newcommand\cN{{\mathcal{N}}}    \newcommand\cR{{\mathcal{R}}}       \newcommand\cY{{\mathcal{Y}}} 

\begin{document}
\title{Rate-Matching Deep Polar Codes\\ via Polar Coded Extension}

\author{   
Geon Choi, \IEEEmembership{Student Member, IEEE} and
 Namyoon Lee, \IEEEmembership{Senior Member, IEEE}
% \thanks{
% This work was supported in part by}
\thanks{
Geon Choi is with the School of Electrical Engineering, Korea University, Seoul 02841, South Korea (e-mail: \mbox{simon03062@korea.ac.kr}).
}
\thanks{Namyoon Lee is with the Department of Electrical Engineering, Pohang University of Science and Technology (POSTECH), Pohang 37673, South Korea (e-mail: \mbox{nylee@postech.ac.kr}).
}
}

\maketitle

\begin{abstract}
  Deep polar codes are pre-transformed polar codes that employ a multi-layered polar kernel transformation strategy to enhance code performance in short blocklength regimes. However, like conventional polar codes, their block length is constrained to powers of two, as the final transformation layer uses a conventional polar kernel matrix. This paper introduces a novel rate-matching technique for deep polar codes using code extension, particularly effective when the desired code length slightly exceeds a power of two. The key idea is to exploit the layered structure of deep polar codes by concatenating polar codewords generated at each transformation layer. Based on this structure, we also develop an efficient decoding algorithm leveraging soft-output successive cancellation list decoding and provide comprehensive error probability analysis supporting our code design algorithms. Additionally, we propose a computationally efficient greedy algorithm for multi-layer configurations. Extensive simulations confirm that our approach delivers substantial coding gains over conventional rate-matching methods, especially in medium to high code-rate regimes.
\end{abstract}

\section{Introduction}
 Polar codes, introduced by Ar{\i}kan, represent a breakthrough in coding theory as the first class of channel codes proven to achieve the capacity of symmetric binary-input memoryless channels under low-complexity successive cancellation (SC) decoding \cite{arikan09-polar}. Their elegant theoretical properties and efficient implementation have led to their adoptation in modern communication systems, most notably in the 5G New Radio (NR) standard \cite{3gpp-nr-coding}. However, the performance of conventional polar codes degrades significantly in the short blocklength regime, which is critical for latency-sensitive applications such as ultra-reliable low-latency communication (URLLC) \cite{shirvanimoghaddam19, yue23, stephan-welcome-chance,BOSS-URLLC,BOSS3}. This degradation stems from suboptimal distance properties, including low minimum distance and many low-weight codewords \cite{mondelli-urbanke-polar2RM}.

 To address these limitations, pre-transformed polar codes have emerged as an enhanced coding scheme for short blocklengths \cite{Niu12-CA-polar, Trifonov-polar-dynamic-frozen, Wang-PCC-polar, Zhang-PC-polar-Huawei, arikan-pac, deep-polar-tcom, spp-tcom, deep-polar-gc-wkshps, spp-isit}. These codes apply an upper-triangular transformation to the input bits before polar encoding, reshaping the input space and improving the code’s minimum distance and weight spectrum. This approach enhances performance under both maximum likelihood (ML) and successive cancellation list (SCL) decoding, particularly with large list sizes.

 Despite these advances, pre-transformed polar codes—like their conventional counterparts—are restricted to blocklengths that are powers of two. Supporting arbitrary code lengths and rates requires rate-matching techniques such as puncturing, shortening, or extension, which are widely used in conventional polar codes \cite{Shin13-puncturing, Kai13-U-rep, Niu13-QUP, Wang14-shortening, Miloslavskaya15-shortening, Saber15-Factor-extension, Zhao-Polar-matrix-extension, Jang19-rate-matching-binary-domination, Jang20-structural-extension}. Puncturing approaches \cite{Shin13-puncturing, Kai13-U-rep, Niu13-QUP, Jang19-rate-matching-binary-domination} such as quasi-uniform puncturing (QUP) remove specific bits from a more extended mother code. Shortening techniques, on the other hand, fix certain bits to known values before transmission \cite{Wang14-shortening, Miloslavskaya15-shortening, Jang19-rate-matching-binary-domination}. Both methods have been thoroughly explored, with optimization frameworks established for different channel conditions and code parameters. These methods perform well under specific rate regimes—puncturing is typically effective for low-rate codes, while shortening is preferred for high-rate scenarios \cite{3gpp-nr-coding}. However, both approaches often introduce additional decoding complexity due to longer mother codes.

 In contrast, when the desired code length exceeds a power of two by a small margin, extension-based methods can offer superior performance with reduced complexity compared to puncturing and shortening \cite{Jang20-structural-extension}. Despite this advantage, extension techniques for polar codes have received comparatively less attention in the literature. The work \cite{Kai13-U-rep, Saber15-Factor-extension, Zhao-Polar-matrix-extension, Jang20-structural-extension} proposed an incremental redundancy hybrid automatic repeat request (HARQ) scheme that extends polar codes for retransmission. Structural extensions using simplex kernels are proposed for rate-matching techniques with a theoretical justification \cite{Jang20-structural-extension}. Nevertheless, the literature on extension-based rate-matching methods designed explicitly for pre-transformed polar codes remains notably sparse, with most existing approaches focusing on conventional polar codes.

 The key challenge in designing effective extension methods lies in maintaining the polarization structure that gives polar codes their desirable properties while effectively utilizing the additional code bits. In this paper, we propose a novel rate-matching framework for deep polar codes, a family of pre-transformed polar codes that enhances short blocklength performance through multi-layered polar kernel transformations \cite{deep-polar-tcom, spp-tcom}. While deep polar codes offer improved error correction, their blocklengths remain restricted to powers of two due to the final polar transformation stage. To address this, we introduce a systematic extension method that leverages the hierarchical encoding structure of deep polar codes by concatenating partial codewords from different transformation layers. This approach maintains the code's structural advantages while enabling flexible blocklengths. Additionally, we develop an efficient soft-output SCL decoding algorithm with soft-in soft-out (SISO) processing to boost performance with minimal complexity overhead.

 Our contributions are summarized as follows:
\begin{itemize}
    \item 
    We develop a novel extension framework for deep polar codes by leveraging their multi-layer hierarchical structure \cite{deep-polar-tcom, spp-tcom}. Our approach utilizes the output of pre-transforms as additional polar codewords, concatenating them with the main codeword to achieve flexible blocklengths. This method effectively exploits the coding gain inherent in the pre-transform layers while maintaining the structural advantages of deep polar codes. 
    \item 
    We design an efficient decoding algorithm incorporating soft information from pre-transform layers into the primary decoding process. By utilizing soft-output SCL decoding \cite{Gao24-SoSCL}, our method exploits future frozen constraints within the SC decoding framework, allowing each layer's component to contribute reliability improvement (akin to deep polarization) that enhances overall decoding performance with minimal complexity overhead.
    \item 
    We establish a comprehensive theoretical analysis framework based on density evolution under Gaussian approximation (DEGA) in \cite{Mori-density-evolution-LCOM, Trifnov-polar-construction, Wu-DEGA-SC-Pe} that incorporates the effects of SISO decoding. This framework enables the optimization of design parameters for both two-layered and multi-layered deep polar code configurations. Additionally, we propose a computationally efficient greedy algorithm that reduces the design complexity from exponential to linear in the number of layers while maintaining reasonable performance.
    \item 
    We validate our approach through extensive simulations across diverse code parameters. Results demonstrate that our proposed extension method significantly outperforms conventional rate-matching techniques, particularly in medium to high code-rate regimes. The notable coding gains and reduced computational complexity make our approach well-suited for practical applications requiring short blocklength codes, such as URLLC and short-packet communications.
\end{itemize}

 The remainder of this paper is organized as follows. Section~II provides preliminaries on polar codes and SISO decoding. Section~III details our proposed single-layer extension method, including encoding, decoding, and design considerations. Section~IV extends these concepts to multi-layer configurations with a greedy approach for rate profiling. Section~V presents comprehensive simulation results that demonstrate the performance advantages of our approach across various code parameters. Finally, Section~VI concludes the paper with a summary of our contributions.

\section{Preliminaries}
\subsection{System Model}
A binary linear block code $\cC(N,K)$ with a codeword length $N$ and code dimension $K$ is defined by the row space of the generator matrix ${\bf G}\in \mathbb{F}_2^{K \times N}$, or equivalently, by the null space of the parity-check matrix ${\bf H} \in \mathbb{F}_2^{(N-K) \times N}$. 
A codeword ${\bf c} = [c_0, c_1, \ldots, c_{N-1}] \in \mathbb{F}_2^N$ is generated from the information block ${\bf m} = [m_0, m_1, \ldots, m_{K-1}] \in \mathbb{F}_2^K$ via the relation ${\bf c} = {\bf m} {\bf G}$, and modulated using binary phase shift keying (BPSK) as ${\bf x} = 1 - 2 {\bf c}$. 
The modulated symbol vector ${\bf x}$ is transmitted over an additive white Gaussian noise (AWGN) channel, resulting in  the received signal ${\bf y} = {\bf x} + {\bf w}$, where the additive noise ${\bf w} = [w_0, w_1, \ldots, w_{N-1}] \in \mathbb{R}^N$ is an independent and identically distributed (i.i.d.) zero-mean Gaussian random variable with variance $\sigma^2 = N_0 / 2$ where $N_0$ is the noise power spectrum density, i.e., $\cN(0, \sigma^2)$.

\subsection{Polar Codes}\label{sec:polar-code}
A polar code with parameters $(N,K,\mathcal{I})$ is characterized by the polar transform matrix of size $N=2^n ~ (n\in \mathbb{N})$ and an index set $\mathcal{I}\subseteq [0,N-1]$ with $|\cI| = K$. The polar transform matrix of size $N=2^n$ is obtained through the $n$th Kronecker power of a binary kernel matrix ${\bf F}_2=\begin{bmatrix}
1 & 0\\
1 & 1 
\end{bmatrix}$ as ${\bf F}_N = {\bf F}_2^{\otimes n}$. 
The generator matrix is the sub-matrix of ${\bf F}_N$, consisting of the rows indexed by $\cI$. 

\subsubsection{Encoding}
By utilizing the structure of ${\bf F}_N$, the low-complexity encoding can be performed. Given a message ${\bf m} \in \mathbb{F}_2^K$, the input of polar transform ${\bf u} \in \mathbb{F}_2^N$ is generated based on the information index set $\cI$ as ${\bf u}_{\cI} := {\bf m}$ and ${\bf u}_{\cI^c} := {\bf 0}$. Here, $\mathcal{I}^c=[0, N-1] \backslash \mathcal{I}$ is referred to as the frozen index set. Then, a polar codeword ${\bf c} \in \mathbb{F}_2^N$ is constructed by applying polar transform to ${\bf u}$. 

\subsubsection{Channel polarization}
A channel polarization is the underlying principle for the design of information index set $\cI$ and the SC decoding \cite{arikan09-polar}. Suppose a modulated codeword ${\bf x}$ is transmitted over binary input discrete output symmetric memoryless channel (B-DMC) $W^N({\bf y}|{\bf x}) \defeq \prod_{i=0}^{N-1} W(y_i|x_i)$, where $y \in \cY$. Define synthetic channels $W_N^{(i)}$ for $i=0, \ldots, N-1$ as 
\begin{align}
   W_N^{(i)}\left({\bf y},  {{\bf u}_{0:i-1} | u_i}\right) = \sum_{{\bf u}_{i+1:N-1} \in \mathbb{F}_2^{N-i-1}} \frac{1}{2^{N-1}} W^N\left({\bf y}  | {\bf x}\right),
\end{align}
where ${\bf x}$ is encoded codeword using ${\bf u}$, and ${\bf u}_{a:b}=[u_a,u_{a+1},\ldots, u_{b}]$ for $a,b\in [N]$ and $a<b$. Note that the symmetric capacity of a B-DMC $W$ is given by 
\begin{align}
    I(W) \defeq \sum_{y \in \cY} \sum_{x \in \mathbb{F}_2} \frac{1}{2}W(y|x)\log\frac{W(y|x)}{\frac{1}{2}W(y|0)+\frac{1}{2}W(y|1)}.
\end{align}
The channel polarization principle states that the symmetric capacities of $W_N^{(i)}$ are polarized into two states as $N$ becomes large, i.e., $I\left(W_N^{(i)}\right) \rightarrow 0$ or $I\left(W_N^{(i)}\right) \rightarrow 1$ as $N\rightarrow \infty$, and the fraction of indices $i$ such that $I\left(W_N^{(i)}\right) \rightarrow 1$ approaches $I(W)$. 
In general, the information index set can be designed as the $K$ best indices in terms of symmetric capacities. 

\subsubsection{SC decoding}
The SC decoder estimates $u_i$ sequentially from $i=0$ to $i=N-1$. If $i \in \cI$, the decoder uses the previous decisions $\hat{\bf u}_{0:i-1}$, and computes the log-likelihood ratio (LLR)
\begin{align}
    L_N^{(i)}({\bf y}, \hat{\bf u}_{0:i-1}) \defeq \log \frac{W_N^{(i)}({\bf y}, {\bf u}_{0:i-1}|0)}{W_N^{(i)}({\bf y}, {\bf u}_{0:i-1}|1)}, \label{eqn:synthetic-llr}
\end{align}
and generates its decision as
% \begin{align}
%     \hat{u}_i = \begin{cases} 0, & \text{if } L_N^{(i)}({\bf y}, \hat{\bf u}_{0:i-1}) > 0, \\ 1, &\text{otherwise}. \end{cases} \label{eqn:sc-decision}
% \end{align}
\begin{align}
    \hat{u}_i = \begin{cases} 0, & \text{if } L_N^{(i)}({\bf y}, \hat{\bf u}_{0:i-1}) > 0, \\ 
    1, & \text{if } L_N^{(i)}({\bf y}, \hat{\bf u}_{0:i-1}) < 0, \\
    0 \text{ or } 1, &\text{if } L_N^{(i)}({\bf y}, \hat{\bf u}_{0:i-1}) = 0. \end{cases} \label{eqn:sc-decision}
\end{align}
If $i\in \cI^c$, the decoder sets the known value $\hat{u}_i := u_i$. Due to the channel polarization, the error probability of estimating $\hat{u}_i$ under correct previous decisions ${\bf u}_{0:i-1}$ for $i \in \cI$ is close to $0$, and the decoding error probability is low enough to achieve the channel capacity \cite{arikan09-polar}.

\begin{figure}
\centering
\input{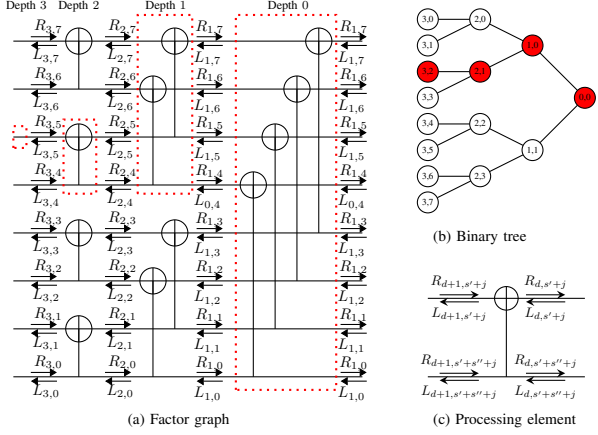}
\caption{Factor graph of polar transform and its corresponding binary tree, whose nodes consist of processing elements.}
\label{fig:polar-code}
\end{figure}

The binary tree representation of the polar transform in Fig.~\ref{fig:polar-code} is convenient to understand the SC decoding process. For each node at depth $d \in [0, n]$, there are $2^{d}$ nodes and we label the $m$th node as $(d,m)$. For a given depth $d$, let $s = n-d$. Each non-leaf node $(d,m)$ node consists of $2^{s-1}$ processing elements, each of size $2 \times 2$, as shown in Fig.~\ref{fig:polar-code}. Define $s' = 2^{s}m$ and $s'' = 2^{s-1}$. Then, the $j$th processing element at detph $d$ consists of  following components: left-to-right messages $(L_{d, s' + j}$, $L_{d, s' + s'' + j})$, right-to-left messages $(R_{d, s' + j}$, $R_{d, s'+ s'' + j})$, and hard decision bits $(U_{d, s' + j}$, $U_{d, s' + s'' + j})$.

The SC decoding process can be visualized as a {\it pre-order} traversal of a binary tree \cite{Niu14-primary-concepts}. Upon visiting a non-leaf node $(d,m)$, the messages are updated according to the following rules:
\begin{align}
    L_{d+1, s' + j} &= f(L_{d, s' + j}, L_{d, s' + s'' + j}), \label{eqn:l2r-first}\\
    L_{d+1, s' + s'' + j} &= g(L_{d, s' + j}, L_{d, s' + s'' + j}, U_{d+1, s' + j}), \label{eqn:l2r-second}\\
    U_{d, s' + j} &= U_{d+1, s' + j} \oplus U_{d+1, s' + s'' + j}, \label{eqn:l2r-third-1} \\
    U_{d, s' + s'' + j} &= U_{d+1, s' + s'' + j}, \label{eqn:l2r-third-2}
\end{align}
where $f(x,y) = \log\left(\frac{1+e^{x+y}}{e^x + e^y} \right)$ and $g(x, y, u) = (1-2u)x + y$. 
During the first visit, the rule in \eqref{eqn:l2r-first} is applied. In the second visit, the rule in \eqref{eqn:l2r-second} is applied. During the third visit, the rules in \eqref{eqn:l2r-third-1} and \eqref{eqn:l2r-third-2} are applied. At a leaf node, the decision for $\hat{u}_i$ is made according to \eqref{eqn:sc-decision}. The initialization is $L_{0, i} = \frac{2y_i}{\sigma^2}$
% At a leaf node is visited, the decision for $\hat{u}_i$ is made according to \eqref{eqn:sc-decision}, where the synthetic LLR is $L_{d, i}$. 

\subsubsection{SCL decoding \cite{Tal15-polar-SCL, Niu12-CA-polar, Stimming15-LLR-SCL}}
In the application of rule \eqref{eqn:l2r-second}, preceding decisions $\hat {\bf u}_{0:i-1}$ are used. To mitigate error propagation, the SCL decoder retains up to $L$ decoding candidate, by evaluating path reliability. The path metric of the $\ell$th decoding candidate is computed at each leaf node by \cite[Th. 1]{Stimming15-LLR-SCL}, 
\begin{align}
    {\sf PM}_i[\ell] \defeq \sum_{j = 0}^{i} \log \left( 1 + e^{-(1 - 2\hat{u}_j [\ell]) \cdot L_{n,j}[\ell] } \right),
\end{align}
where $L_{n,j}[\ell]$ is leaf node message computed at the $\ell$th decoding path with preceding decisions $\hat{\bf u}_{0:j-1}[\ell]$. The path metric represents the path reliability by the relation, 
\begin{align}
    {\sf PM}_i[\ell] = -\log \mathbb{P}({\bf U}_{0:i} = \hat{\bf u}_{0:i}[\ell]  \vert {\bf Y} = {\bf y}).
\end{align}
At each leaf node, the decoder selects the $L$ decoding candidates with the smallest path metrics.

\subsubsection{SISO decoding \cite{Gao24-SoSCL}}
The standard SCL decoder belongs to soft-in hard-out decoder. Obtaining soft-output is important for our proposed extension method. Several algorithms can be used to compute the soft output for $x_i$, defined as 
\begin{align}
    {\Lambda}_i \defeq \frac{\mathbb{P}(x_i = +1 | {\bf y})}{\mathbb{P}(x_i = -1|{\bf y})}.
\end{align}
In this work, we employ the SCL decoding-based method presented in \cite{Gao24-SoSCL}, referred to as {\it SoSCL} in the sequel. It is equivalent to the {\it post-order} traversal of the binary tree, during which the following rules are applied:
\begin{align}
    & R_{d, s'+j} = f(R_{d+1, s'+j}, R_{d+1, s'+s''+j} + L_{d, s'+s''+j}), \label{eqn:l2r-1} \\
    & R_{d, s'+s''+j} = f(R_{d+1, s'+j}, L_{d, s'+j}) + R_{d+1, s'+s''+j}. \label{eqn:l2r-2}
\end{align} 
Note that we omit the decoding path index $\ell$ for convenience. Each $R_{d,j}$ is computed for each decoding candidates $\ell$ with different $L_{d, j}$ due to different $U_{d, j}$. 
The initialization is $R_{n,i} = \infty$ if $i\in \cI^c$ and $R_{n,i} = 0$ if $i \in \cI$.

Following the traversal process, the soft outputs $\Lambda_i$ are obtained by aggregating the traversal outputs $R_{0,i}[\ell]$ across all $\ell$ decoding paths. As demonstrated in \cite{Gao24-SoSCL}, each individual soft output $R_{0,i}[\ell]$ adheres to the following relationship:
\begin{align}
    \Lambda_i[\ell] &= \log \frac{ \mathbb{P}(x_i = +1 \vert {\bf Y} = {\bf y}, {\bf U} = \hat{\bf u}[\ell]) } { \mathbb{P}(x_i = -1 \vert {\bf Y} = {\bf y}, {\bf U} = \hat{\bf u}[\ell] ) } \\
    & \approx L_{0,i} + R_{0,i}[\ell],
\end{align}
where $\Lambda_i[\ell]$ represents the SCL soft-output corresponding to the $\ell$th candidate decoding path $\hat{\bf u}[\ell]$. The precision of this approximation is contingent upon the specific algorithm employed to compute $R_{0,i}[\ell]$. 
The final soft output $\Lambda_i$ can then be obtained by performing a weighted combination of all $R_{0,i}[\ell]$ values using path metrics ${\sf PM}_N[\ell] = \mathbb{P}({\bf U} = \hat{\bf u}[\ell] | {\bf Y} = {\bf y})$ as weighting factors.

% Then, the final soft output $\Lambda_i$ can be calculated, by following the chain of \eqref{eqn:soft-output-1} to \eqref{eqn:soft-output-2}, where $\cC$ is the set of all the valid codewords, and $\cL$ is the decoding candidates of the SCL decoder at the last stage. 

\subsection{Rate-Matching}
 The length of polar codes derived from Ar{\i}kan's $2 \times 2$ polarization kernel is fixed to a power of $2$, i.e., $N = 2^n$. Rate-matching methods such as puncturing, shortening, repetition, and extension can be used to create codewords of arbitrary length, denoted as $M$. Puncturing or shortening is used to reduce the length of a code, while extension (including repetition) is employed to increase code length. Puncturing and shortening start from a larger mother polar code, whereas repetition and extension begin with a smaller polar code. A representative puncturing and shortening method is puncturing the first $N-M$ codeword bits, or shortening the last $N-M$ codeword bits, often referred to as the {\it quasi uniform} manner. For the 5G NR rate-matching, the quasi-uniform puncturing and shortening with sub-block interleaving is employed.

There is no single rate-matching method that consistently outperforms others across all scenarios. The decoding performance of modified polar codes is influenced by the codeword length $M$ and code dimension $K$. Typically, puncturing yields better results for low-rate codes, while shortening is more effective for high-rate codes. However, both techniques require decoding a larger mother code, which increases computational complexity. In some instances, code extension offers a more favorable trade-off between performance and complexity, particularly when the target code length slightly exceeds a power of two. In such scenarios—especially at low code rates—repetition or extension can outperform both puncturing and shortening.

In this work, we focus on the extension approach, specifically for cases where the desired code length slightly exceeds a power of two. Although repetition is generally preferred for low-rate codes, our proposed scheme, based on extended deep polar codes, demonstrates superior performance for medium and high-rate regimes.

\section{Single-Layer Extended Deep Polar Codes}
This section presents the encoding and decoding methods for rate-matched deep polar codes, focusing on two-layer architectures for clarity. The extension to multi-layer designs shall be explained in Section~\ref{sec:multi-layer}.
% N-layered deep polar codes -> (N-1)-layer extended deep polar codes. 

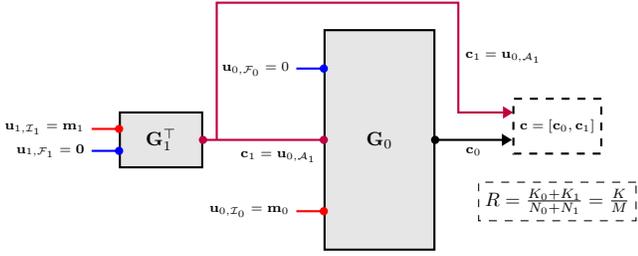
\begin{figure}
    \centering
    \usetikzlibrary{calc, arrows.meta, positioning, fit}
\begin{tikzpicture}[
    scale=0.73,
    transform shape,
    >={Triangle[scale=0.8]},
    thick,
    block/.style={rectangle, draw, anchor=south west, fill=gray!20, thick},
    dot/.style={circle, minimum size=0.15cm, inner sep=0pt},
]

% \draw[help lines] (-4,-4) grid (6,6);

% Blocks
\node[block, minimum height=4cm, minimum width=2cm] (FN) {$\mathbf{G}_{0}$};
\node[block, minimum height=1cm, minimum width=1.5cm, left= 2.2cm of FN] (FN1) {$\mathbf{G}_{1}^\top$};

% Input signals for FN1
\coordinate (FN1_info) at ($(FN1.west)+(0,0.2)$);
\coordinate (FN1_frozen) at ($(FN1.west)-(0,0.2)$);

\node[dot, fill=red] at (FN1_info) {};
\node[dot, fill=blue] at (FN1_frozen) {};
\draw[red, thick] (FN1_info) -- ++(-0.5,0) coordinate (FN1_info_text);
\draw[blue, thick] (FN1_frozen) -- ++(-0.5,0) coordinate (FN1_frozen_text);
\node[anchor=east, font=\scriptsize] at (FN1_info_text) {$\mathbf{u}_{1,\mathcal{I}_1} = {\bf m}_1$};
\node[anchor=east, font=\scriptsize] at (FN1_frozen_text) {$\mathbf{u}_{1,\mathcal{F}_1} = {\bf 0}$};

% Output signals from FN1 to FN
\coordinate (FN1_out) at ($(FN1.east)$);
\coordinate (FN_connection) at ($(FN.west)$);
\coordinate (FN_frozen) at ($(FN.west) + (0,1.3)$);
\coordinate (FN_info) at ($(FN.west) - (0,1.3)$);
\coordinate (FN_out) at ($(FN.east)$);

\node[dot, fill=purple] at (FN1_out) {};
\draw[purple, thick] (FN1_out) -- (FN_connection);
\node[dot, fill=purple] at (FN_connection) {};
\node[anchor=east, font=\scriptsize] at ($(FN_connection)+(0, -0.3)$) {${\bf c}_1 = {\bf u}_{0,\mathcal{A}_1}$};

% Input signals for FN
\node[dot, fill=red] at (FN_info) {};
\node[dot, fill=blue] at (FN_frozen) {};
\draw[blue, thick] (FN_frozen) -- ++(-0.5,0) coordinate (FN_frozen_text);
\draw[red, thick] (FN_info) -- ++(-0.5,0) coordinate (FN_info_text);
\node[anchor=east, font=\scriptsize] at (FN_frozen_text){$\mathbf{u}_{0,\mathcal{F}_0} = 0$};
\node[anchor=east, font=\scriptsize] at (FN_info_text) {$\mathbf{u}_{0,\mathcal{I}_0} = {\bf m}_0$};

\node[dot, fill=black] at (FN_out) {};
\draw[black, thick, ->] (FN_out) -- ++(1.4, 0) node[below, pos=0.5, font=\scriptsize]{${\bf c}_0$} coordinate (CWD_main);
\draw[purple, thick, ->] ($(FN1_out)+(0.25, 0)$) -- ++(0,2.5) -- ++(4.4,0) -- ++(0,-2) node[black, right, pos=0.5, font=\scriptsize]{${\bf c}_1 = {\bf u}_{0,\mathcal{A}_1}$} -- ++(1.0, 0) coordinate (CWD_sub);

\node[anchor=west, draw=black, dashed, minimum height = 1cm, font=\scriptsize] (CWD) at ($(CWD_main)!0.5!(CWD_sub)$) {${\bf c} = \left[ {\bf c}_0, {\bf c}_1 \right]$};

\node[below= 5mm of CWD, draw=black, thin, dashed] {$R = \frac{K_0 + K_1}{N_0 + N_1} =\frac{K}{M}$};

% Message splitter?

\end{tikzpicture}
    \caption{Illustration of the encoding for single-layer extended deep polar codes.}
    \label{fig:single-extension-deep-polar-enc}
\end{figure}

\subsection{Encoding}
Our code extension method leverages the structure of successive encoding of deep polar codes \cite{deep-polar-tcom,spp-tcom}.  Fig.~\ref{fig:single-extension-deep-polar-enc} illustrates the two-stage successive encoding process of a two-layered deep polar code. 
The generator matrices for the layer 0 and layer 1 of the deep polar codes with size of $N_0=2^{n_0}$ and $N_1=2^{n_1}$ are a submatrix of ${\bf G}_i$ defined as
\begin{align}
    {\bf G}_i = \mathbf{F}_2^{\otimes n_i},
\end{align}
 where $i\in \{0,1\}$ and \(\otimes\) denotes the Kronecker (tensor) product. As can be seen in Fig.~\ref{fig:single-extension-deep-polar-enc}, unlike PAC-like codes, our deep polar codes uses the pre-transform matrix by multiplying the transpose of the polar transform matrix  ${\bf G}_{1}^{\top} \in \mathbb{F}_2^{N_1\times N_1}$.

Let \({\bf m}_i \in \mathbb{F}_2^{K_i}\) denote the information message for encoding layer \(i\), where each message has length \(K_i\) for \(i \in \{0, 1\}\). We define the information bit set \(\mathcal{I}_i \subset [N_i]\) and the frozen bit set \(\mathcal{F}_i \subset [N_i]\) corresponding to layer \(i\). Additionally, let \(\mathcal{A}_1 \subset [N_0]\) represent the connection bit set for layer 0.  For layer 0, the information, connection, and frozen bit sets are mutually exclusive and collectively exhaustive, satisfying \(\mathcal{I}_0 \cap \mathcal{A}_1 = \mathcal{I}_0 \cap \mathcal{F}_0 = \mathcal{A}_1 \cap \mathcal{F}_0 = \emptyset\) and \(\mathcal{I}_0 \cup \mathcal{A}_1 \cup \mathcal{F}_0 = [N_0]\).  For layer 1, the information and frozen bit sets are disjoint and together span the full code length, i.e., \(\mathcal{I}_1 \cap \mathcal{F}_1 = \emptyset\) and \(\mathcal{I}_1 \cup \mathcal{F}_1 = [N_1]\).

\subsubsection{Encoding for Layer 1}  
The information message ${\bf m}_1 \in \mathbb{F}_2^{K_1}$ is mapped to the information bit positions indexed by \(\mathcal{I}_1\), resulting in the vector \({\bf u}_{1,\mathcal{I}_1}\). The remaining positions corresponding to the frozen set \(\mathcal{F}_1\) are filled with zeros, denoted as \({\bf u}_{1,\mathcal{F}_1} := \mathbf{0} \in \mathbb{F}_2^{N_1 - K_1}\). The full input vector to the polar transform is thus \([{\bf u}_{1,\mathcal{I}_1},~{\bf u}_{1,\mathcal{F}_1}]\), and the encoded codeword for layer 1 is given by\footnote{For readability, we omit the permutation operation from our notation as it would only increase notational complexity unnecessarily.}
\begin{align}
    {\bf c}_1 = [{\bf u}_{1,\mathcal{I}_1},~{\bf u}_{1,\mathcal{F}_1}] \cdot {\bf G}_1^{\top} \in \mathbb{F}_2^{N_1},
\end{align}
where \({\bf G}_1\) is the polar transform matrix of length \(N_1\).

\subsubsection{Encoding for Layer 0}
For layer 0, the message vector \({\bf m}_0 \in \mathbb{F}_2^{K_0}\) is mapped into the positions defined by the information set \(\mathcal{I}_0\), forming \({\bf u}_{0,\mathcal{I}_0}\). Additionally, the encoded output from layer 1, denoted \({\bf c}_1 \in \mathbb{F}_2^{N_1}\), is embedded into the connection set \(\mathcal{A}_1\), resulting in \({\bf u}_{0,\mathcal{A}_1} := {\bf c}_1\). The frozen bits are placed in the positions defined by \(\mathcal{F}_0\), represented as \({\bf u}_{0,\mathcal{F}_0} := \mathbf{0} \in \mathbb{F}_2^{N_0 - K_0 - N_1}\). The final codeword for layer 0 is computed as
\begin{align}
    {\bf c}_0 = [{\bf u}_{0,\mathcal{I}_0},~{\bf u}_{0,\mathcal{A}_1},~{\bf u}_{0,\mathcal{F}_0}] \cdot {\bf G}_0 \in \mathbb{F}_2^{N_0},
\end{align}
where \({\bf G}_0\) is the polar transform matrix for length \(N_0\).

\subsubsection{Code Concatenation for Extension}
The deep polar code construction generates two codewords: \({\bf c}_0 \in \mathbb{F}_2^{N_0}\) from layer 0 and \({\bf c}_1 \in \mathbb{F}_2^{N_1}\) from layer 1. Our extension strategy is to concatenate these codewords to form a longer code:
\begin{align}
    {\bf c} = [{\bf c}_0,~{\bf c}_1] \in \mathbb{F}_2^{N_0 + N_1}.
\end{align}
The overall code rate of the extended code is given by
\begin{align}
    R = \frac{K_0 + K_1}{N_0 + N_1}=\frac{K}{M}.
\end{align}

This concatenation-based extension enables flexible blocklengths that are not limited to powers of two, as \(N_0 + N_1\) does not need to satisfy \(2^n\) for any integer \(n\). In addition, by carefully designing the information sets \(\mathcal{I}_0\) and \(\mathcal{I}_1\), the proposed framework can support a wide range of code rates, making it suitable for practical scenarios requiring rate adaptivity and flexible blocklengths.

To construct the proposed extension code, it is crucial to properly select the sets $(\cI_0, \cI_1, \cA_1)$ for each encoding layer. In section~\ref{sec:degignSPP}, we shall present a method for designing these sets based on block error rate (BLER) analysis based on DEGA \cite{Mori-density-evolution-LCOM, Trifnov-polar-construction, Wu-DEGA-SC-Pe}.

\begin{figure}
    \centering
    \begin{tikzpicture}[
        >={Triangle[scale=0.8]},
        thick,
        block/.style={rectangle, draw, anchor=south west, fill=none, thick, inner sep=2mm},
        dot/.style={rectangle, minimum size=1mm, inner sep=0pt},
        every node/.style={font=\footnotesize}
    ]
    % \draw [help lines] grid(4,4);
    
    \node (y) {${\bf y} = \begin{bmatrix} {\bf y}_1 \\ {\bf y}_0 \end{bmatrix}$};
    \coordinate (y_sub) at ($(y.east)+(0,0.2)$);
    \coordinate (y_main) at ($(y.east)+(0,-0.2)$);
    \node[dot, fill=black] at (y_sub) {};
    \node[dot, fill=black] at (y_main) {};

    \node[block, above=12mm of y_sub, text width=14mm, align=flush center] (decode_sub) {SoSCL Decoding};

    \draw[black, ->] (y_sub) -- coordinate[pos=0.4](y_sub_label) (decode_sub.south);
    \node[right=0mm of y_sub_label] {${\bf L}_1 = \frac{2{\bf y}_1}{\sigma^2}$};

    \node[block, right=30mm of y_main, text width=14mm, align=flush center] (decode_main) {Modified SCL Decoding};
    \draw[black, ->] (y_main) -- coordinate[pos=0.5](y_main_label) (decode_main.west);
    \path (decode_sub.east) edge[black, ->, bend left=30] coordinate[pos=0.5](soft_out_label) (decode_main.north);

    \node[right=3mm of soft_out_label] {$\Lambda_i$ (to depth $n$)};
    \node[below=0mm of y_main_label, text width = 15mm] {${\bf L}_0 = \frac{2{\bf y}_0}{\sigma^2}$ (to depth 0)};

    \draw[black, ->] (decode_main.east) -- ++(1,0) coordinate[pos=0.5] (decode_out_label);
    \node[black, above=0mm of decode_out_label] {$\hat{\bf u}_0$};
    
    \end{tikzpicture}
    \caption{The proposed LLR combined decoding process.}
    \label{fig:single-extension-deep-polar-dec}
\end{figure}
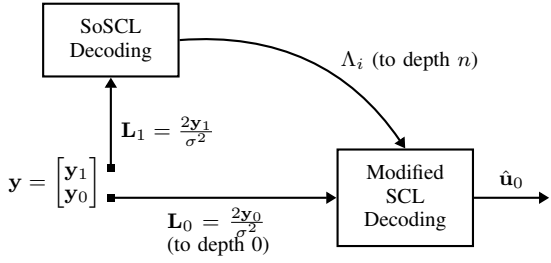

\subsection{LLR Combined SCL Decoding}
The decoder exploits the structure of the proposed code, which consists of two locally SCL-decodable codewords, ${\bf c}_0$ and ${\bf c}_1$. Notably, ${\bf c}_1$ is a transposed version of a standard polar codeword, resulting in a reversed bit order compared to conventional polar encoding. Let ${\bf y} = [{\bf y}_0, {\bf y}_1]$ denote the received signal vector from the channel, where ${\bf y}_0$ and ${\bf y}_1$ are noisy versions of modulated codewords ${\bf x}_0$ and ${\bf x}_1$, respectively. The decoding process is illustrated in Fig.~\ref{fig:single-extension-deep-polar-dec}.

The decoder first computes the LLR vector ${\bf L}_1$ using ${\bf y}_1$ and estimates the soft information associated with the connection bits ${\bf u}_{0,\mathcal{A}_1} = {\bf c}_1$, which are shared between the two layers of the deep polar code. The LLR for the $i$-th connection bit is computed as:
\begin{align}
    \Lambda_{i}^{(1)} &= \log \frac{\mathbb{P}(c_{1,i} = 0 \mid {\bf y}_1)}{\mathbb{P}(c_{1,i} = 1 \mid {\bf y}_1)} \\
    &= \log \frac{\mathbb{P}(x_{1,i} = +1 \mid {\bf y}_1)}{\mathbb{P}(x_{1,i} = -1 \mid {\bf y}_1)},
\end{align}
where $x_{1,i}$ denotes the modulated symbol corresponding to the $i$-th code bit $c_{1,i}$.

For convenience, we construct an \textit{embedded soft information vector} $\boldsymbol{\Lambda} = [\Lambda_0, \ldots, \Lambda_{N_0-1}] \in \mathbb{R}^{N_0}$, where the entries are assigned as follows: for each connection index $\mathcal{A}_{1,j}$, the corresponding value is set to $\Lambda_{\mathcal{A}_{1,i}} = \Lambda_i^{(1)}$; for all remaining indices, the entries are set to zero, i.e., $\Lambda_i = 0$. Here, $\mathcal{A}_{1,i}$ denotes the $i$-th element in the set $\mathcal{A}_1$, assumed to be in natural ascending order.

Subsequently, the deep polar code is decoded using a LLR-combined SCL decoder that incorporates both the LLR vector ${\bf L}_0$ corresponding to the received signal ${\bf y}_0$ and the soft information $\Lambda_i$ associated with the connection bits ${\bf u}_{0, \mathcal{A}_1}$. The decoder estimates the transmitted message $\hat{\bf m}$—or equivalently, the input vector $\hat{\bf u}_0$—from which the original message is recovered via inverse pre-transformation.

The key modification in the SCL decoder lies in the combination of the soft information $\Lambda_i$—obtained from the SoSCL decoder for ${\bf y}_1$—with the original LLR values $L_{n,i}$ in the SCL decoder for ${\bf y}_0$. The combined LLR used in decoding is defined as
\begin{align}
    \tilde{L}_{i} = \begin{cases}
        \Lambda_i + L_{n, i}, & \text{if } i \in \mathcal{A}_{1, \mathcal{I}_1}, \\
        L_{n, i}, & \text{otherwise},
    \end{cases}
    % \label{eqn:soft-combining}
\end{align}
where $\mathcal{A}_{1,\mathcal{I}_1} = \{a_i : i \in \mathcal{I}_1\}$ denotes the indices in the connection set $\mathcal{A}_1$ that correspond to information bits.

Using this modified LLR $\tilde{L}_i$, the path metric for each decoding path $\ell$ is updated according to
\begin{align}
    \mathsf{PM}_i[\ell] = \mathsf{PM}_{i-1}[\ell] + \log\left(1 + e^{-(1 - 2\hat{u}_i[\ell]) \cdot \tilde{L}_i} \right),
    \label{eqn:modified-PM}
\end{align}
which ensures that the influence of the soft information is reflected during the path extension process.

\vspace{0.2cm}
\begin{remark}[Effect of Soft Information $\Lambda_i$]
    In a conventional SCL decoder, frozen bits following the current bit are treated as uniformly random (i.e., unknown), and LLRs are used without incorporating side information. As a result, when decoding the first connection bit $\hat{u}_{\mathcal{A}_{1,1}}$, the decoder assumes that all subsequent connection bits $\hat{u}_{\mathcal{A}_{1,1}^c}$ are information bits. However, with SoSCL decoding applied to ${\bf y}_1$, the soft information vector $\Lambda$ encodes partial reliability knowledge, including for subsequent frozen bits indexed by $\mathcal{A}_{1,\mathcal{F}_1}$. This enhances the reliability of decoding $\hat{u}_{\mathcal{A}_{1,1}}$ by incorporating soft constraints on its context.
\end{remark}

\subsection{Decoding Error Probability Analysis}\label{sec:single-error-probability}

We present a decoding error probability analysis under SC decoding. Analyzing the decoding error probability based on the exact density evolution of LLR values is generally intractable, as it requires characterizing the full distribution of LLRs, which becomes analytically complex. To overcome this difficulty, we adopt the DEGA approach \cite{Trifnov-polar-construction, Wu-DEGA-SC-Pe}. This method approximates the distribution of each LLR value \(L_{d,i}\) as a Gaussian random variable with mean \(\mu_{d,i}\) and variance \(\sigma^2_{d,i}\), i.e., 
\[
L_{d,i} \sim \mathcal{N}(\mu_{d,i}, \sigma^2_{d,i}).
\]
Due to the symmetry of the underlying binary-input memoryless channel, the variance of the LLR is determined by its mean through the relation \(\sigma^2_{d,i} = 2\mu_{d,i}\). This approximation greatly simplifies the analysis while still providing accurate estimates of the decoding error probability under SC decoding.

Following the same notation as in \eqref{eqn:l2r-first}–\eqref{eqn:l2r-third-2}, the mean value \(\mu_{d,i}\) of the LLR at depth \(d\) under SC decoding can be estimated using the DEGA method. Specifically, the mean updates follow the recursive equations:
\begin{align}
    &\mathbb{E}\left[ \tanh\left( \frac{L_{d+1, s'+j}}{2} \right) \right] \nonumber \\
    &= \mathbb{E}\left[ \tanh\left( \frac{L_{d, s'+j}}{2} \right) \right] \cdot \mathbb{E}\left[ \tanh\left( \frac{L_{d, s'+s''+j}}{2} \right) \right], \label{eqn:DEGA-L1}
\end{align}
and
\begin{align}
    \mu_{d+1, s'+s''+j} = \mu_{d, s'+j} + \mu_{d, s'+s''+j}, \label{eqn:DEGA-L2}
\end{align}
where \(L_{d,i}\) denotes the LLR at depth \(d\) and index \(i\), and \(s', s''\) define the recursive structure of the polar transform.

The identity in \eqref{eqn:DEGA-L1} can alternatively be expressed in terms of the function \(\psi(\cdot)\), defined as 
% $\psi(\mu) = \mathbb{E}\left[ \tanh\left( \frac{X}{2} \right) \right]$
\[
\psi(\mu) = \mathbb{E}\left[ \tanh\left( \frac{X}{2} \right) \right],
\]
where \(X \sim \mathcal{N}(\mu, 2\mu)\). Then, the mean update becomes:
\begin{align}
    \psi(\mu_{d+1, s'+j}) = \psi(\mu_{d, s'+j}) \cdot \psi(\mu_{d, s'+s''+j}).
\end{align}
The initial mean value is given by \(\mu_{0,i} = \frac{2}{\sigma^2}\), where \(\sigma^2\) is the noise variance of the channel. This recursive formulation allows efficient estimation of LLR means for all bit indices under the Gaussian approximation and zero-codeword transmission.

Note that the decision of $\hat{u}_i$ uses the value of $\tilde{L}_{n,i} = L_{n,i} + \mathds{1}[i \in \cA_{1, \cI_1}]\Lambda_i$. To estimate $\mathbb{E}[\tilde{L}_{n,i}]$, the mean value of soft information $\Lambda_i$, or $R_{0,i}$, is required. 
To address this, we propose applying the same approach, though $R_{n,i}$ is a binary value and no longer satisfies the Gaussian assumption. Specifically, we assume that $R_{d,i} \sim \cN(\eta_{d,i}, 2\eta_{d,i})$ and apply the following rules:
\begin{align}
    & \psi(\eta_{d, s'+j}) \nonumber \\ & \quad = \psi(\eta_{d+1, s'+j}) \psi(\eta_{d+1, s'+s''+j} + \mu_{d, s'+s''+j}), \label{eqn:DEGA-R1}\\
    & \eta_{d, s'+s''+j} \nonumber \\
    & \quad = \eta_{d+1, s'+s''+j} + \psi^{-1} \left( \psi(\eta_{d+1, s'+j}) \psi (\mu_{d, s'+j})\right). \label{eqn:DEGA-R2}
\end{align}
Here, the initial value is $\eta_{n, i} = 0$ if $i \in \cI$ and $\eta_{n,i} = \infty$ if $i \in \cF$. Note that we overload the notation of $\mu_{n,j}$; depending on the context, it refers to the mean of LLR $L_{n,j}$ corresponding to either ${\bf y}_1$ or ${\bf y}_0$.

% It is worth noting that \eqref{eqn:DEGA-L1} and \eqref{eqn:DEGA-L2} are applied for ${\bf x}_0$, while \eqref{eqn:DEGA-R1} and \eqref{eqn:DEGA-R2} are applied for ${\bf x}_1$. 

The SC decoding error probability \( {\sf P}_{e,0} \) for the codeword corresponding to \({\bf x}_0\) can be approximated following \cite{Trifnov-polar-construction, Wu-DEGA-SC-Pe} with a slight modification to include $\eta_{0,i}$ as
\begin{align}
    {\sf P}_{e,0} &= 1 - \prod_{i \in \mathcal{I}_0} \left( 1 - Q\left( \sqrt{\frac{\mu_{n,i}}{2}} \right) \right) \nonumber \\
    &\quad\quad \times \prod_{i \in \mathcal{I}_1} \left( 1 - Q\left( \sqrt{\frac{\mu_{n, \mathcal{A}_{1,i}} + \eta_{0,N_1-i}}{2}} \right) \right), 
    \label{eqn:P-error-v1}
\end{align}
where \( \eta_{0,N_1-i} \) denotes the soft information passed from the decoding of layer 1 and the index $N_1-i$ arises from the vector reversal induced by the transpose operation.

Note that the expressions in \eqref{eqn:DEGA-R1} and \eqref{eqn:DEGA-R2} assume a genie-aided scenario in which each bit \( U_i \) is decoded using the correct application of the update rule \eqref{eqn:DEGA-L2}, implying perfect side information. As a result, the estimate in \eqref{eqn:P-error-v1} may overestimate the reliability of the soft output term \( \eta_{0,i} \), and thus provides an optimistic prediction of the overall error probability.

To obtain a more conservative and practical estimate for code design, we account for the possibility that decoding of layer 1 may fail. Let \( {\sf P}_{e,1} \) denote the SC decoding error probability for \({\bf x}_1\). Then, the total decoding error probability \( {\sf P}_e \) can be upper bounded as
\begin{align}
    {\sf P}_e \lesssim 1 - (1 - {\sf P}_{e,1})(1 - {\sf P}_{e,0}),
    \label{eqn:P-error-v2}
\end{align}
which reflects the dependency between decoding failures in the two layers. In the following sections, we adopt \eqref{eqn:P-error-v2} as the primary metric for the design of deep polar codes.

\begin{remark}
We provide two expressions for the decoding error probability: \eqref{eqn:P-error-v1} and \eqref{eqn:P-error-v2}. When the deep polar code is well-designed and decoding of \({\bf x}_1\) is successful, \eqref{eqn:P-error-v1} tends to be more accurate, as illustrated in Fig.~\ref{fig:BLER-M1088}. However, this expression may be overly optimistic due to its assumption of perfectly reliable soft output \(\eta_{0,i}\), especially when \(K_1\) is large.  In contrast, \eqref{eqn:P-error-v2} assumes that the soft output is completely unreliable if the decoding of \({\bf y}_1\) fails, resulting in a more conservative estimate. While pessimistic, it offers a more reliable performance prediction as a function of \(K_1\). Therefore, we adopt \eqref{eqn:P-error-v2} as the primary design metric in Section~\ref{sec:degignSPP}.
\end{remark}

\subsection{Rate-Profiling Method for Extended Deep Polar Codes} \label{sec:degignSPP}

Designing a deep polar code involves determining the pre-transform length \(N_1\) and selecting the information and connection index sets \((\mathcal{I}_0, \mathcal{I}_1, \mathcal{A}_1)\). The design criteria should satisfy the following objectives:
\begin{enumerate}
    \item The decoding of \({\bf c}_1\) must be successful in order to obtain reliable soft information \(\Lambda_i\), and
    \item The bit channels with soft information, characterized by \(\mu_{n,\mathcal{A}_{1,i}} + \eta_{0,N_1-i}\), should exhibit high reliability to ensure correct decoding of \({\bf c}_0\).
\end{enumerate}

To enable successful decoding of \({\bf c}_1\) using  SoSCL, the parameters \(\mathcal{I}_1\), \(K_1 = |\mathcal{I}_1|\), and \(N_1\) must be carefully selected. In particular, the code rate \(K_1/N_1\) should not exceed the channel capacity, and the set \(\mathcal{I}_1\) should be chosen based on a bit channel reliability metric such as symmetric capacity or estimated LLR means.

For decoding \({\bf c}_0\), the sets \(\mathcal{I}_0\) and \(\mathcal{A}_1\) play a central role. The decoding reliability of bit \(u_i\) is determined by the synthesized channel \(W({\bf y}_0, {\bf u}_{0:i-1} | u_i)\). For binary-input AWGN channels, the quality of this channel can be approximated via density evolution of the LLR mean \(\mu_{n,i}\), possibly using Gaussian approximation. A larger \(\mu_{n,i}\) indicates a lower error probability for bit \(i\), assuming correct decoding of all prior bits.

For each index \(j \in \mathcal{A}_1\) that is connected to an information bit \(i \in \mathcal{I}_1\), i.e., \(j = \mathcal{A}_{1,i}\), the reliability is enhanced by soft information, making the effective LLR mean \(\mu_{n,j} + \eta_{0,N_1-i}\). Consequently, a practical approach is to select \(\mathcal{I}_0\) as the \(K_1\) most reliable indices, while choosing \(\mathcal{A}_1\) as the subsequent \(N_1\) best indices based on \(\mu_{n,i}\). 
Then, the dimension of pre-transform $K_1$ is optimized to minimize \eqref{eqn:P-error-v2}.

\subsection{Example}

\begin{figure}
    \centering
    \includegraphics[width=\columnwidth]{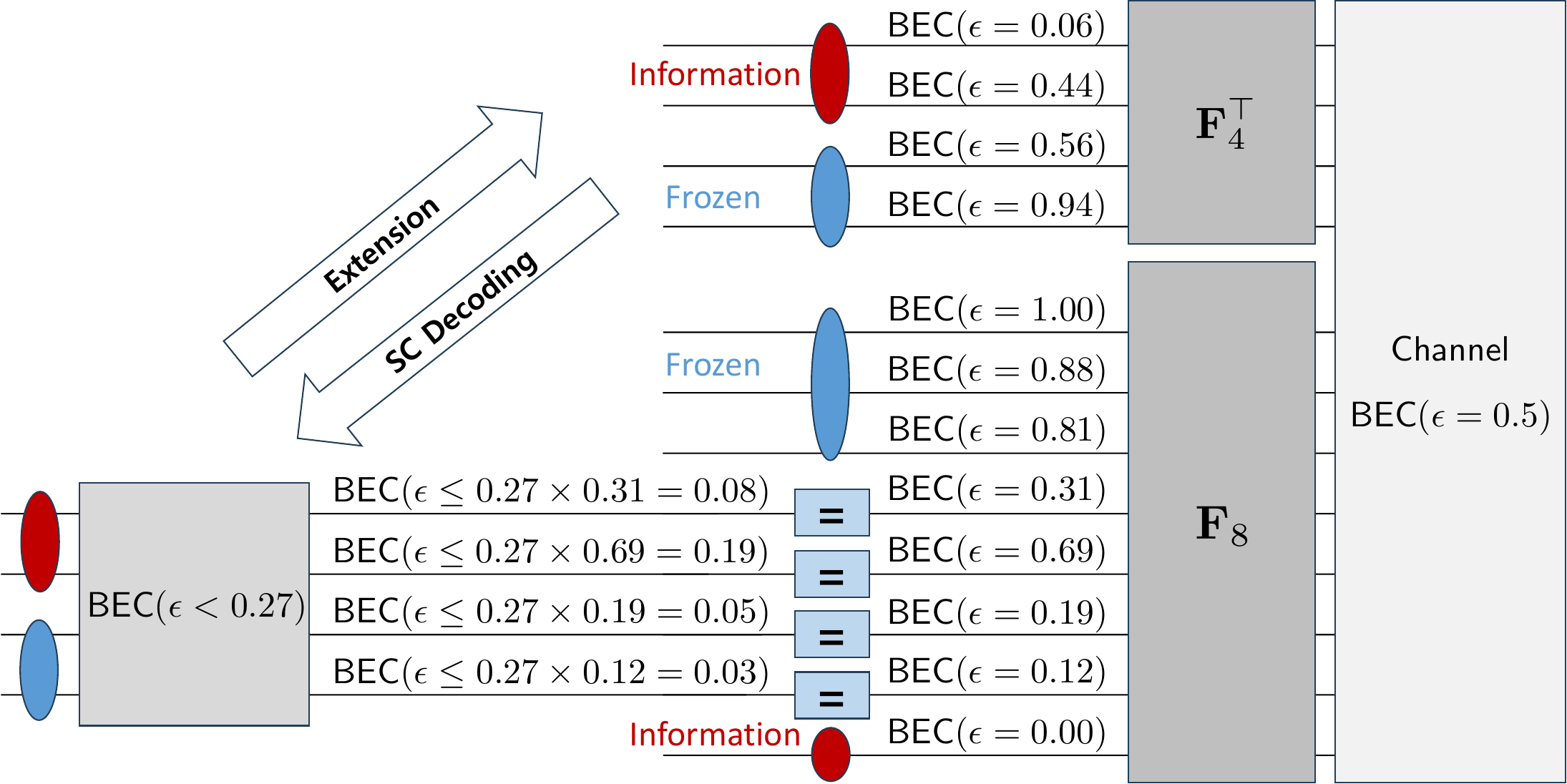}    
    \caption{An example of extended deep polar codes with parameters $(N_0, N_1, K)=(8,4,3)$ under a BEC with an erasure probability of $\epsilon=0.5$.}
    \label{fig:example}
\end{figure}

We present an example for $M=12$ and $K=3$ over a binary erasure channel (BEC) to demonstrate how extended deep polar codes utilize additional extended bits to improve code performance under SC decoding. 
Fig.~\ref{fig:example} illustrates the scenario under consideration. We construct a deep polar code with parameters $N_0 = 8$ and $N_1 = 4$ and extend its pre-transformed output ${\bf c}_1$ computed by matrix ${\bf F}_4^\top$. Since the extended part ${\bf c}_1$ is transmitted through a BEC with erasure probability $\epsilon = 0.5$, and it consists of polar codewords, the polarized bit channels are BEC with erasure probability $\epsilon \in \{0.06, 0.44, 0.56, 0.94\}$. Suppose that we transmit information through the two best-polarized channels. Then, under SC decoding, the decoding is successful if: i) both $u_3$ and $u_4$ are correct, or ii) $u_3$ is erased but correctly guessed with probability $0.5$ and $u_4$ is correct. Although the erased $u_4$ can be guessed with probability $0.5$, we ignore these cases and treat our BLER as an upper bound. The corresponding BLER is upper-bounded by 
\begin{align}
    {\sf BLER} &< 1- (1-0.44)(1-0.06) - \frac{1}{2}\times 0.44\times(1-0.06) \nonumber \\
    & = \frac{27}{100}.
\end{align}
After decoding of the extended part, the corresponding pre-transformed parts can be regarded as transmitted through BEC with erasure probability less than $0.27$. In our extended deep polar codes, this information is combined with the original polarized channels through an equality constraint, resulting in upgraded BECs with erasure probability $0.27 \times \epsilon_i$, where $\epsilon_i$ is the erasure probability of the $i$th polarized channel. Without extension, the information is carried through the bit channels with $\epsilon_i \in \{0.19, 0.12, 0.00\}$, which are improved to $\epsilon \in \{0.08, 0.19, 0.00\}$.

% \subsection{Latency, Memory, Complexity Analysis}

\section{Multi-Layer Extended Deep Polar Codes}\label{sec:multi-layer}

In this section, we present multi-layered deep polar codes to construct extended codes of length $M < 2N$ with information dimension $K < N$, where $N = 2^n$ is the original polar code length.

\subsection{Encoding}

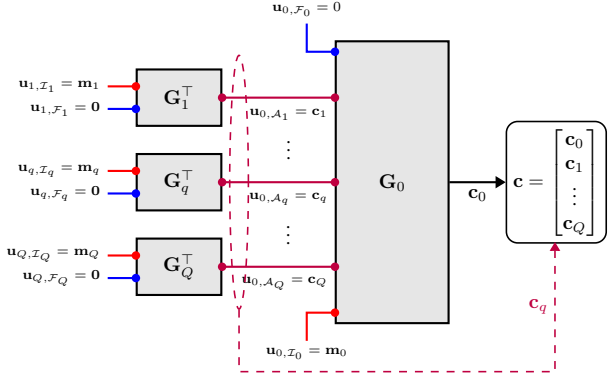
\begin{figure}
    \centering
    \usetikzlibrary{calc, arrows.meta, positioning, fit, shapes.geometric}
\begin{tikzpicture}[
    scale=0.75,
    transform shape,
    >={Triangle[scale=0.8]},
    thick,
    block/.style={rectangle, draw, anchor=south west, fill=gray!20, thick},
    dot/.style={circle, minimum size=0.15cm, inner sep=0pt},
]

% \draw[help lines] (-4,-4) grid (6,6);

\pgfmathsetmacro{\FNOneHeight}{1.5}
\pgfmathsetmacro{\FNTwoHeight}{0}
\pgfmathsetmacro{\FNThreeHeight}{-1.5}

% Blocks
\node[block, minimum height=5cm, minimum width=2cm] (FN) {$\mathbf{G}_{0}$};
\node[block, anchor=east, minimum height=1cm, minimum width=1.5cm] at ($(FN.west) + (-2, \FNOneHeight)$) (FN1) {$\mathbf{G}_{1}^\top$};
\node[block, anchor=east, minimum height=1cm, minimum width=1.5cm] (FN2) at ($(FN.west) + (-2, \FNTwoHeight)$) {$\mathbf{G}_{q}^\top$};
\node[block, anchor=east, minimum height=1cm, minimum width=1.5cm] at ($(FN.west) + (-2, \FNThreeHeight)$)(FN3) {${\bf G}_Q^\top$};

% Input signals for FN1
\coordinate (FN1_info) at ($(FN1.west)+(0,0.2)$);
\coordinate (FN1_frozen) at ($(FN1.west)-(0,0.2)$);

\node[dot, fill=red] at (FN1_info) {};
\node[dot, fill=blue] at (FN1_frozen) {};
\draw[red, thick] (FN1_info) -- ++(-0.5,0) coordinate (FN1_info_text);
\draw[blue, thick] (FN1_frozen) -- ++(-0.5,0) coordinate (FN1_frozen_text);
\node[anchor=east, font=\scriptsize] at (FN1_info_text) {$\mathbf{u}_{1,\mathcal{I}_1} = {\bf m}_1$};
\node[anchor=east, font=\scriptsize] at (FN1_frozen_text) {$\mathbf{u}_{1,\mathcal{F}_1} = {\bf 0}$};

% Output signals from FN1 to FN
\coordinate (FN1_out) at ($(FN1.east)$);
\coordinate (FN_connection1) at ($(FN.west) + (0, \FNOneHeight)$);

\node[dot, fill=purple] at (FN1_out) {};
\node[dot, fill=purple] at (FN_connection1) {};
\draw[purple, thick] (FN1_out) -- (FN_connection1);
\node[anchor=east, font=\scriptsize] (u1_text) at ($(FN_connection1)+(0, -0.3)$) {$\mathbf{u}_{0,\mathcal{A}_1} = {\bf c}_1$};

\node at ($(u1_text.south) + (0, -0.3)$) {$\vdots$};

% Input signals for FN2
\coordinate (FN2_info) at ($(FN2.west)+(0,0.2)$);
\coordinate (FN2_frozen) at ($(FN2.west)-(0,0.2)$);

\node[dot, fill=red] at (FN2_info) {};
\node[dot, fill=blue] at (FN2_frozen) {};
\draw[red, thick] (FN2_info) -- ++(-0.5,0) coordinate (FN2_info_text);
\draw[blue, thick] (FN2_frozen) -- ++(-0.5,0) coordinate (FN2_frozen_text);
\node[anchor=east, font=\scriptsize] at (FN2_info_text) {$\mathbf{u}_{q,\mathcal{I}_q} = {\bf m}_q$};
\node[anchor=east, font=\scriptsize] at (FN2_frozen_text) {$\mathbf{u}_{q,\mathcal{F}_q} = {\bf 0}$};

% Output signals from FN2 to FN
\coordinate (FN2_out) at ($(FN2.east)$);
\coordinate (FN_connection2) at ($(FN.west) + (0, \FNTwoHeight)$);

\node[dot, fill=purple] at (FN2_out) {};
\node[dot, fill=purple] at (FN_connection2) {};
\draw[purple, thick] (FN2_out) -- (FN_connection2);
\node[anchor=east, font=\scriptsize] (u2_text) at ($(FN_connection2)+(0, -0.3)$) {$\mathbf{u}_{0,\mathcal{A}_q} = {\bf c}_q$};

\node at ($(u2_text.south) + (0, -0.3)$) {$\vdots$};

% Input signals for FN3
\coordinate (FN3_info) at ($(FN3.west)+(0,0.2)$);
\coordinate (FN3_frozen) at ($(FN3.west)-(0,0.2)$);

\node[dot, fill=red] at (FN3_info) {};
\node[dot, fill=blue] at (FN3_frozen) {};
\draw[red, thick] (FN3_info) -- ++(-0.5,0) coordinate (FN3_info_text);
\draw[blue, thick] (FN3_frozen) -- ++(-0.5,0) coordinate (FN3_frozen_text);
\node[anchor=east, font=\scriptsize] at (FN3_info_text) {$\mathbf{u}_{Q,\mathcal{I}_Q} = {\bf m}_Q$};
\node[anchor=east, font=\scriptsize] at (FN3_frozen_text) {$\mathbf{u}_{Q,\mathcal{F}_Q} = {\bf 0}$};

% Output signals from FN3 to FN
\coordinate (FN3_out) at ($(FN3.east)$);
\coordinate (FN_connection3) at ($(FN.west) + (0, \FNThreeHeight)$);

\node[dot, fill=purple] at (FN3_out) {};
\node[dot, fill=purple] at (FN_connection3) {};
\draw[purple, thick] (FN3_out) -- (FN_connection3);
\node[anchor=east, font=\scriptsize] at ($(FN_connection3)+(0, -0.3)$) {$\mathbf{u}_{0,\mathcal{A}_Q} = {\bf c}_Q$};

% Input signals for FN
\coordinate (FN_frozen) at ($(FN.north west) - (0,0.2)$);
\coordinate (FN_info) at ($(FN.south west) + (0,0.2)$);
\coordinate (FN_out) at ($(FN.east)$);

\node[dot, fill=red] at (FN_info) {};
\node[dot, fill=blue] at (FN_frozen) {};
\draw[blue, thick] (FN_frozen) -- ++(-0.5,0) -- ++ (0, 0.5) coordinate (FN_frozen_text);
\draw[red, thick] (FN_info) -- ++(-0.5,0) -- ++ (0, -0.5) coordinate (FN_info_text);
\node[anchor=south, font=\scriptsize] at (FN_frozen_text){$\mathbf{u}_{0,\mathcal{F}_0} = 0$};
\node[anchor=north, font=\scriptsize] at (FN_info_text) {$\mathbf{u}_{0,\mathcal{I}_0} = {\bf m}_0$};

% Output signals for FN
\draw[black, thick, ->] (FN_out) -- ++(1.0, 0) node[below, pos=0.5]{${\bf c}_0$} coordinate (CWD_main);

% Codeword
\node[anchor=west, draw, minimum height = 1cm, rounded corners=.15cm, semithick] (CWD) at ($(CWD_main)$) {${\bf c} = \begin{bmatrix} {\bf c}_0 \\ {\bf c}_1 \\ \vdots \\ {\bf c}_Q \end{bmatrix}$};

% connection indices ellipse
\node[ellipse, draw=purple, minimum width = 0.2cm, 
    minimum height = 4.5cm, semithick, dashed] (con) at ($(FN2.east) + (0.3, 0)$) {};
\draw[purple, semithick, dashed, ->] ($(con.south)$) -- ++(0, -1.1) -- ++(5.6,0) -- ++(0,2.3) 
% node[left, pos=0.7]{${\bf c}_1 = {\bf u}_1$} 
node[left, pos=0.5]{${\bf c}_q$};
% node[left, pos=0.3]{${\bf c}_Q = {\bf u}_Q$};

% \node[below= 5mm of CWD, draw=black, thin, dashed] {$R = \frac{K}{N_1 + N}$};

% Message splitter?

\end{tikzpicture}
    \caption{The encoding for multi-layer extended deep polar codes.}
    \label{fig:multi-extension-deep-polar-enc}
\end{figure}

Fig.~\ref{fig:multi-extension-deep-polar-enc} illustrates the encoding process of a $(Q+1)$-layered deep polar code. 
The generator matrices for each layer $q$ of deep polar codes with size of $N_q= 2^{n_q}$ are a submatrix of ${\bf }G_q$ defined as 
\begin{align}
    {\bf G}_q = {\bf F}_2^{\otimes n_q}.
\end{align}
As can be seen in Fig.~\ref{fig:multi-extension-deep-polar-enc}, our deep polar codes use the pre-transform matrix by multiplying the transpose of the polar transform matrix ${\bf G}_q^\top$ in parallel. 

Analogous to the two-layered deep polar codes, the information message ${\bf m}$ splits into $Q+1$ sub-vector ${\bf m}_i \in \mathbb{F}_2^{K_i}$. For $1\le q\le Q$, each sub-vector ${\bf m}_q$ is mapped to the information bit positions indexed by information bit set $\cI_q \in [N_q]$, resulting in ${\bf u}_{q, \cI_q} := {\bf m}_q$. The remaining positions are filled with zeros, denoted as ${\bf u}_{q, \cF_q} := {\bf 0}$ where $\cF_q = [N_q] \backslash \cI_q$. The full input vector $[{\bf u}_{q, \cI_q}, {\bf u}_{q, \cF_q}]$ is transformed by the polar transform matrix ${\bf G}_q^\top$, and the encoded codeword for layer $q$ is given by
\begin{align}
    {\bf c}_q = [{\bf u}_{q, \cI_q}, {\bf u}_{q, \cF_q}] \cdot {\bf G}_q^\top \in \mathbb{F}_2^{N_q},
\end{align}
which is then assigned to the $q$th connection index set $\cA_q$, resulting in ${\bf u}_{0, \cA_q} := {\bf c}_q$. Note that ${\bf c}_q$ is a polar codeword with size $N_q$. 
With the input vector of layer 0 assigned as ${\bf u}_{0, \cI_0} := {\bf m}_0$, ${\bf u}_{0, \cF_0} := {\bf 0}$, and ${\bf u}_{0, \cA_q} := {\bf c}_q$ for $1\le q \le Q$, the deep polar codeword is computed as
\begin{align}
    {\bf c}_0 = [{\bf u}_{0, \cI_0}, {\bf u}_{0, \cA_1}, \ldots, {\bf u}_{0, \cA_Q}, {\bf u}_{0, \cF_0}] \cdot {\bf G}_0 \in \mathbb{F}_2^{N_0}.
\end{align}

The $(Q+1)$-layered deep polar code generates $Q+1$ codewords ${\bf c}_q$ with size $N_q$. Similar to single-layer extension, we concatenate all these codewords to form a longer code as 
\begin{align}
    {\bf c} = [{\bf c}_0, {\bf c}_1, {\bf c}_2, \ldots, {\bf c}_Q] \in \mathbb{F}_2^{N_0 + N_1 + \cdots + N_Q},
\end{align}
resulting in $Q$-layer extended codeword length of $M = \sum_{q=0}^Q N_q$ and code rate of $R = K / M$.

\begin{remark}[Parameters of deep polar codes given $M$]
    The size of the pre-transform $N_\ell$ for multi-layered deep polar codes is determined by the extension size $M-N$.   By representing $M - N = \sum_{q = 0}^{\log_2(N)-1} l_{q} 2^{q}$ in binary form, where $l_q \in \{0, 1\}$, we derive a natural deep polar code construction in which the number of pre-transforms is $Q = \sum_{q=0}^{\log_2(N)-1} l_q$ and the size of the $q$-th pre-transform is $N_q = 2^q$. While this construction offers a systematic approach to non-power-of-two code lengths, it should be noted that it may not yield optimal decoding performance. Nevertheless, we adopt this approach in this paper to design extended deep polar codes for arbitrary code length $M$.
\end{remark}

\subsection{Decoding}
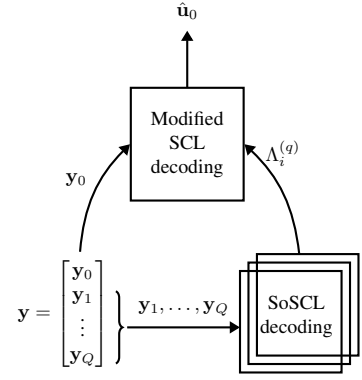
\begin{figure}
    \centering
    \usetikzlibrary{calc, arrows.meta, positioning, fit, shapes.geometric, decorations.pathreplacing}

\begin{tikzpicture}[
        scale=0.75,
        transform shape,
        >={Triangle[scale=0.8]},
        thick,
        block/.style={rectangle, draw, fill=none, thick, inner sep=2mm},
        dot/.style={rectangle, minimum size=1mm, inner sep=0pt},
        % every node/.style={font=\footnotesize}
    ]

    % \draw [help lines] grid(4,4);
    
    \node (y) {${\bf y} = \begin{bmatrix} {\bf y}_0 \\ {\bf y}_1 \\ \vdots \\ {\bf y}_Q \end{bmatrix}$};

    \draw[decoration={brace}, decorate] ($(y.north east) + (0, -0.7)$) -- node[right=-1pt] {} ($(y.south east) + (0, 0.2)$) coordinate[pos=0.5] (y_sub);

    \draw[black, ->] ($(y_sub) + (0.2, 0)$) --  coordinate[pos=0.5](y_sub_label)  ++(2,0) coordinate (SoSCL);
    % \draw[black, ->] ($(y_sub) + (0.3, 0.1)$) -- ++(2,0);
    % \draw[black, ->] ($(y_sub) + (0.4, 0.2)$) --  coordinate[pos=0.5](y_sub_label) ++(2,0) ;
    \node[above= 2pt of y_sub_label] {${\bf y}_1, \dots, {\bf y}_Q$};

    \draw ($(SoSCL) + (0, -0.8)$) rectangle ($(SoSCL) + (1.8,1)$);
    \draw ($(SoSCL) + (0.15, -0.65)$) rectangle ($(SoSCL) + (1.95,1.15)$);
    \draw ($(SoSCL) + (0.30, -0.50)$) rectangle ($(SoSCL) + (2.10,1.30)$);
    
    \node[align=center] at ($(SoSCL) + (1,0.2)$) {SoSCL \\ decoding};

    \node[draw, align=center, minimum width = 2cm, minimum height = 2cm] (SCL) at (2.2,3) {Modified \\ SCL \\ decoding};

    \draw[->] ($(SoSCL) + (1.05,1.3)$) to[bend right=20] node[pos=0.5, above=10pt, align=center]{$\Lambda_i^{(q)}$} ($(SCL.east)$) ;

    \draw[->] ($(y.north)+(0.3,0)$) to [bend left=20] node[pos=0.5, auto, align=center] {${\bf y}_0$} ($(SCL.west)$);    

    \draw[->] ($(SCL.north)$) -- ++(0,1) node[above=2pt]{$\hat{\bf u}_0$};

\end{tikzpicture}
    \caption{An illustration of the decoding method for multi-layer extended deep polar codes.}
    \label{fig:multi-extension-dec}
\end{figure}

In multi-layered deep polar codes, the multiple sub-codewords ${\bf c}_1, {\bf c}_2, \ldots, {\bf c}_Q$ can be decoded in parallel using polar code decoders. The $Q$ parallel SoSCL decoder uses the received signal vector ${\bf y}_q$ to estimate ${\bf c}_q$ and produce its soft output represented as
\begin{align}
    \Lambda_{i}^{(q)} &= \log \frac{\mathbb{P}(c_{q,i} = 0 \vert {\bf y}_q)}{\mathbb{P}(c_{q,i} = 1 \vert {\bf y}_q)} \\
    &= \log \frac{\mathbb{P}(x_{q,i} = +1 \vert {\bf y}_q)}{\mathbb{P}(x_{q,i} = -1 \vert {\bf y}_q)},
\end{align}
where $x_{q,i}$ denotes the modulated symbol corresponding to the $i$-th code bit $c_{q,i}$ of the $q$th sub-codeword ${\bf c}_q$.

The embedded soft information vector ${\bf \Lambda} = [\Lambda_0,  \ldots, \allowbreak \Lambda_{N_0-1}]$ is defined as $\Lambda_{\cA_{q,j}} = \Lambda_j^{(q)}$ for each connection index, with $\Lambda_{i} = 0$ assigned to the remaining indices.
Subsequently, the modified SCL decoder decodes ${\bf c}_0$ using the received signal ${\bf y}_0$ and the embedded soft information vector ${\bf \Lambda}$. The modified LLR $\tilde{L}_i$ is calculated as
\begin{align}
    \tilde{L}_i = \begin{cases}
        \Lambda_i + L_{n, i}, & i\in\cA_{q, \cI_q} \text{ for some }q, \\
        L_{n,i}, & \text{otherwise}, 
        \end{cases}
        % \label{eqn:soft-combining}
\end{align}
and the path metric is updated according to the procedure described in \eqref{eqn:modified-PM}.

\subsection{Decoding Error Probability}
The computation of SC decoding error probability for each component codeword ${\bf c}_q$ and the average value of soft output follows the DEGA approach as described in Section~\ref{sec:single-error-probability}. Using this method, we can derive the decoding error probability ${\sf P}_{e,0}$ for the extended deep polar code as
\begin{align}
    {\sf P}_{e,0}&= 1 - \prod_{i \in \cI_0} \left( 1 - Q\left( \sqrt{\frac{\mu_{n,i}}{2}}\right) \right) \nonumber \\ 
    &\quad\quad \times \prod_{q = 1}^{Q}\prod_{i \in \cI_q} \left (1 - Q\left( \sqrt{\frac{\mu_{n,\cA_{q,i}} + \eta_{0,N_q-i}^{(q)}}{2}}\right)  \right) \label{eqn:P-error-multi-v1}
\end{align}
where $\eta_{0,i}^{(q)}$ represents the result of density evolution for the $q$th sub-codeword ${\bf c}_q$. 
The modified design metric can be approximated as 
\begin{align}
    {\sf P}_e \lesssim 1-\prod_{q=0}^{Q}(1-{\sf P}_{e,q}), \label{eqn:P-error-multi-v2}
\end{align}
where ${\sf P}_{e, q}$ is the SC decoding error probability for the $q$th sub-codeword ${\bf c}_{q}$.

\subsection{Rate-Profiling}
Rate profiling—the selection of the pre-transform size and information set parameters \((N_q, K_q)\), along with the corresponding information and connection sets \((\mathcal{I}_q, \mathcal{A}_q)\)—is critical for constructing high-performing extended deep polar codes.  
Unfortunately, optimizing these parameters is highly challenging due to the complex interactions between the code structure and decoding performance.  
In this subsection, we present a systematic approach to optimize the code construction parameters efficiently by leveraging the decoding error probability derived in \eqref{eqn:P-error-multi-v2}.

\subsubsection{Comprehensive Iterative Approach}
The process begins with the values of $N_q$ being fixed by the binary representation of the extension size $M-N$, as previously described. To determine the optimal dimensions $K_q$ and the corresponding index sets, we employ an iterative search strategy:
\begin{enumerate}
    \item We explore the space of all possible combinations of $K_q$ values that satisfy the constraint $\sum_{q=0}^{Q} K_q = K$.
    \item For each candidate combination of $K_q$ values, we determine the corresponding index sets $(\mathcal{I}_q, \mathcal{A}_q)$.
    \item For each complete configuration of $K_q$ values and corresponding index sets $(\mathcal{I}_q, \mathcal{A}_q)$, we compute the design metric given in equation~\eqref{eqn:P-error-multi-v2}.
    \item Finally, we select the combination of $K_q$ values that minimizes the design metric as our optimal rate profile.
\end{enumerate}

In the second step of the process, the determination of the index set follows an approach similar to that of single-layer extended deep polar codes, as outlined in Algorithm~\ref{Alg:RP}. The process begins with layer 0, where the $K_0$ most reliable indices are allocated to $\cI_0$, followed by the allocation of the subsequent $N_1$ most reliable indices to $\cA_1$, then the next $N_2$ indices to $\cA_2$, and so forth. To ensure the reliability of the soft information, the information index set for layer $q$, denoted as $\cI_q$, is selected based on reliability order. In the simulation section, we refer to this method as {\it DEGA-UB}.

\begin{algorithm}[t]\label{Alg:RP}
\caption{Design of index sets for deep polar codes}
\KwData{$Q$, $K_0$, $(N_{q})_{q=1}^{Q}$.}
\KwResult{$\{\mathcal{I}_{q}\}_{q=0}^{Q}$, $\{\mathcal{A}_{q}\}_{q=1}^{Q}$.}
\smallskip

$\cR \defeq \{ i_0, i_1, i_2, \ldots, i_{N-1}\}$ // reliability sequence from the most reliable index\;

\medskip
\emph{/* Rate-profile */} \\
$\cI_0 \gets \{i_0, i_1, \ldots, i_{K_0-1}\}$\;

${\sf idx} \gets K_0$\;
\For {$q=1$ \KwTo $Q$}{
    $\mathcal{A}_{q} \gets \{i_{{\sf idx}}, i_{{\sf idx}+1},  i_{{\sf idx}+N_q-1} \}$\;
    ${\sf idx} \gets {\sf idx} + N_q$\;
    $(\cI_q, \cF_q) \gets $ any design methods for polar codes\;
}

$\cF_0 \gets$ remaining indices\;

\end{algorithm}

\subsubsection{Efficient Greedy Approach}
 Rate-profiling for the multi-layer extended deep polar codes differs significantly from the single-layer extension case, where only one parameter $K_1$ needs to be determined. In the multi-layer extension, the search space expands from approximately $N_1$ possible configurations to approximately $N_1 \times N_2 \times \cdots \times N_Q$ configurations, substantially increasing the computational complexity of the design process. This exponential growth in the search space highlights the need for more efficient optimization strategies for practical implementations.

To address this complexity, we propose a greedy-based search outlined in Algorithm~\ref{Alg:find-K-greedy}, which provides an efficient method for determining the optimal rate distribution across multiple layers. The key insight behind this algorithm is to sequentially optimize each layer's dimension parameter, starting from layer $0$ and progressing through to layer $Q-1$, with the final layer's dimension determined by the constraint.

The algorithm considers two rates in each stage: $K_0/N_0$ for the main code and $\bar{K}_1/N_1$ for the first extended part, where $\bar{K}_1 = K - K_0$ represents all remaining information bits. For each potential value of $K_0$, we construct corresponding polar codes and estimate their SC decoding error probability using DEGA. The objective is to find the {\it crossing point} where both polar codes achieve comparable error performance.

Based on empirical observations from our simulation, we implement a specific criterion: we identify the largest value of $K_0$ such that $P_{e,0} < P_{e,1}$, where $P_{e,0}$ and $P_{e,1}$ represent the error probabilities of the respective polar codes. Once the optimal $K_0^\star$ is determined and fixed, we allocate the remaining $K - K_0^\star$ information bits to subsequent layers and repeat the same process iteratively until all layer dimensions are optimized.
This greedy approach reduces the computational complexity from exponential to linear in the number of layers, making it practical for multi-layer extended deep polar code design while still achieving acceptable performance. By considering the error probabilities of two polar codes, the algorithm ensures that information bits are distributed across layers in a manner that balances reliability and optimizes overall error performance.

\begin{algorithm}[t]\label{Alg:find-K-greedy}
\caption{Find $K_q$ (Greedy)}
\KwData{$(N_{q})_{q=0}^{Q}$, $K$}
\KwResult{$(K_{q}^\star)_{q=0}^{Q}$.}

\For {$q=0$ \KwTo $Q-1$} {
    $K_q^\star \gets 0$\;
    \For {$K_q = N_q$ \KwTo $1$} {
        \emph{/* Check for invalid configuration */} \\
        \If{$K_q > \sum_{i=q+1}^{L}N_q$}{{\bf break}\;}
        \If{$q=0$ and $K_0 + \sum_{q=1}^{Q} N_q > N$} {{\bf continue}\;}
        \medskip
        \emph{/* Preceding layer: $\sum_{i = 0}^{q-1}K_i^\star$ */}\\
        \emph{/* Succeeding layer: $(Q-q)$ */}\\
        $\bar{K} \gets {\rm min}(N_{q+1}, K-\sum_{i = 0}^{q-1}K_i^\star - K_q- (Q-q) )$\;
        \medskip
        \emph{/* Find crossing point */} \\
        $P_{e, q} \gets$ Error probability of ${\sf Polar}(N_q, K_q)$\;
        $P_{e, q+1} \gets $ Error probability of ${\sf Polar}(N_{q+1}, \bar{K})$\;
        \If{$P_{e,q} < P_{e, q+1}$}{
        % $K_q^\star \gets K_q$\;
        % ${\sf iscross} \gets {\sf True}$\;
        {\bf break}\;}
    }
    % $K_q^\star \gets {\rm min}(K_q^\star +1, N_q)$\; 
    % \If{${\sf iscross}={\sf False}$} {$K_q^\star \gets 1$\;}
    $K_q^\star \gets$ the largest $K_q$ such that $P_{e,q} < P_{e, q+1}$\; 
}
$K_Q^\star = K - \sum_{q=0}^{Q-1} K_q^\star$\;

\end{algorithm}

\subsection{Discussion on Selection of $N_q$}
Originally, the pre-transform size $N_q$ was determined by the binary representation of the extension size $M-N$. However, alternative deep polar code configurations are also possible. For example, when $N_q = 4$, configurations with $N_q = N_{q+1} = 2$ or $N_q = N_{q+1} = N_{q+2} = N_{q+3} = 1$ are feasible. The proposed decoding method relies on SoSCL decoding of extended  layers, particularly on the assumption that sub-codeword ${\bf c}_q$ is a polar codeword with parameters $(N_q, K_q)$. This implies that the $(N_q, K_q)$ polar code must be successfully decoded; in the event of decoding failure, simple LLR combining would actually provide more reliable information than attempting to use soft outputs.

Avoiding decoding failures of sub-polar codewords is of primary importance. The primary motivation for using soft outputs is to obtain the coding gain of sub-polar codewords. However, if decoding fails, there is no advantage in using the soft-output decoder. In such cases, LLR combining without decoding would be favorable, making it advantageous to use rate-1 codes without parity bits. For this purpose, a variant can be designed that selects $N_q$ information bits with low reliability in vector ${\bf u}$ for repetition. Generally, when $N_q$ is small and soft outputs cannot compensate for the reliability of $u_i$, setting $N_q=1$ is predicted to improve code performance. Verification of this hypothesis remains to be done in future work.

\section{Performance Evaluation}\label{sec:simulation}

We evaluate the decoding performance of CRC-aided (CA) polar codes by presenting the BLER over the binary-input additive white Gaussian noise (BI-AWGN) channel. The simulation setup follows the 5G NR specifications, using the channel-independent reliability sequence and the CRC polynomial
\[
g_{\sf CRC11}(x) = x^{11} + x^{10} + x^9 + x^5 + 1,
\]
as defined in \cite{3gpp-nr-coding}.

% We conducted a comprehensive evaluation of the decoding performance of CRC-aided (CA) polar codes by presenting the BLER over the binary-input additive white Gaussian noise (BI-AWGN) channel. The simulation setup follows the 5G NR specifications, using the channel-independent reliability sequence and the CRC polynomial
% \[
% g_{\sf CRC11}(x) = x^{11} + x^{10} + x^9 + x^5 + 1,
% \]
% as defined in \cite{3gpp-nr-coding}.

For rate-matching operations, we implemented the standardized sub-block interleaver from \cite{3gpp-nr-coding}. Specifically, the following techniques were applied:
\begin{itemize}
    \item \textbf{Puncturing:} The initial \( M - N_0 \) bits of the interleaved codeword are systematically removed.
    \item \textbf{Shortening:} The terminal \( M - N_0 \) interleaved bits are constrained to predetermined values (predominantly zeros) and consequently omitted from transmission.
    \item \textbf{Repetition:} The initial \( M - N_0 \) bits of the interleaved output are duplicated to augment the code length.
\end{itemize}
 This setup ensures consistency with the 5G NR polar coding standard while providing a unified framework for comparative analysis of diverse rate-matching strategies.

For the extended deep polar code design, we implemented the 5G reliability sequence and followed the approach delineated in Algorithm~\ref{Alg:RP}. The parameter $K_q$ was determined through an exhaustive search to optimize \eqref{eqn:P-error-multi-v2} for {\it DEGA-UB} in general cases, while Algorithm~\ref{Alg:find-K-greedy} was specifically applied for the {\it greedy} optimization approach.

It is important to highlight the differentiation in decoder configurations across the evaluated schemes. We assessed code performance under SCL decoding, wherein CRC bits serve for error detection functionality. Puncturing and shortening methods necessitate a larger mother code size relative to repetition and deep polar-based extension techniques. To maintain comparable computational complexity in the asymptotic domain, we employed a doubled list size for repetition and deep polar-based extension relative to puncturing and shortening schemes.

\subsection{BLER Performance}

\begin{figure}
    \centering
    \includegraphics[width=1\columnwidth]{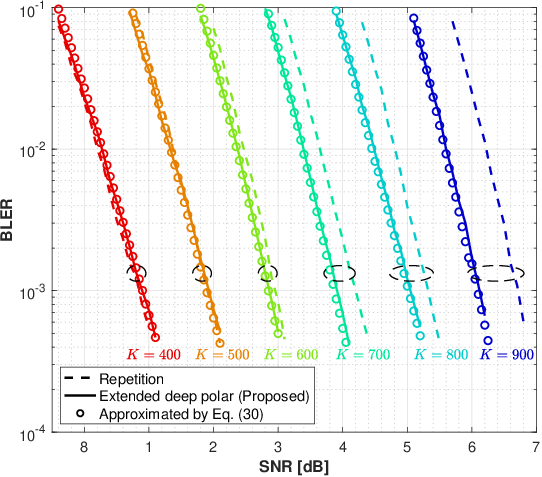}
    \caption{BLER performance of repetition-based and the extended deep polar codes with $N=1024$ and $M=1024+64=1088$, when SC decoding is applied.
    % CRC polynomial $x^{11} + x^{10} + x^{9} + x^{5} + 1$ is used.
    }
    \label{fig:BLER-M1088}
\end{figure}

\begin{figure}
    \centering
    \includegraphics[width=1\columnwidth]{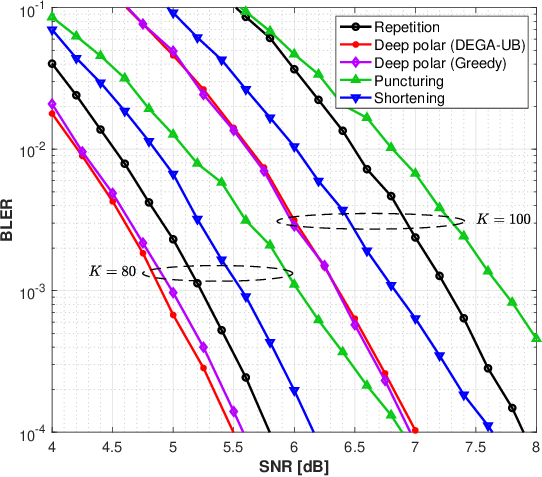}
    \caption{BLER performance of various rate-matched polar codes with $M=152$ and $K\in \{80, 100\}$. The SCL decoding is used with list size $1$ for puncturing and shortening, and list size $2$ for other methods.
    }
    \label{fig:BLER-M152}
\end{figure}

\begin{figure}
    \centering
    \includegraphics[width=1\columnwidth]{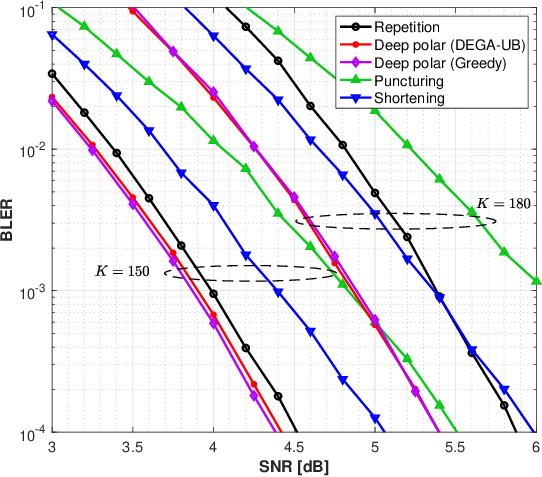}
    \caption{BLER performance of various rate-matched polar codes with $M=304$ and $K\in \{150, 180\}$. The SCL decoding is used with list size $1$ for puncturing and shortening, and list size $2$ for other methods.
    }
    \label{fig:BLER-M304}
\end{figure}

Fig.~\ref{fig:BLER-M1088} illustrates the BLER performance between repetition-based rate-matched polar coding and the extended deep polar code at codeword length $M=1088$. Both coding schemes are constructed from a mother code with dimension $N = 1024$. For the extended deep polar codes, we implemented a single-layer extension configuration with $N_1 = 64$. To maintain error detection capabilities, we deployed SC decoding with an $11$-bit CRC. We investigated a comprehensive range of code rates spanning from moderate $(K=400)$ to high $(K=900)$, where $K$ denotes the message length prior to CRC augmentation. Additionally, we present the theoretical approximation of SC decoding error probability derived from \eqref{eqn:P-error-v1} for single-layer extended deep polar codes, wherein CRC bits are treated as information bits, effectively considering $K+11$ information indices in the calculation. The results conclusively demonstrate that as the code rate increases, the extended deep polar code exhibits significantly superior BLER performance compared to repetition-based rate-matching strategies.

Fig.~\ref{fig:BLER-M152} and \ref{fig:BLER-M304} illustrate the BLER performance of various rate-matched polar codes at code lengths $M=152$ and $M=304$. For each $M\in \{152, 304\}$, the mother code with size $N=\{256, 512\}$ is used for puncturing and shortening, and the mother code with size $N\in \{128, 256\}$ is used for repetition and extended deep polar codes. For decoding, we employed SCL with different list sizes: list size $1$ for puncturing and shortening, and list size $2$ for repetition and extended deep polar codes. The deep polar code implementation uses a three-layer structure where for $M=152$, we utilized $N_1 = 16$ and $N_2 = 8$ ($152 = 128 + 16 + 8)$, and for $M=304$, we employed $N_1 = 32$ and $N_2 = 16$ ($304 = 256 + 32 + 16$). At medium code rates, our proposed methods demonstrate superior performance compared to conventional rate-matching approaches across the examined scenarios.

\subsection{Required SNR}

\begin{figure*}
    \centering
    \includegraphics[width=\textwidth]{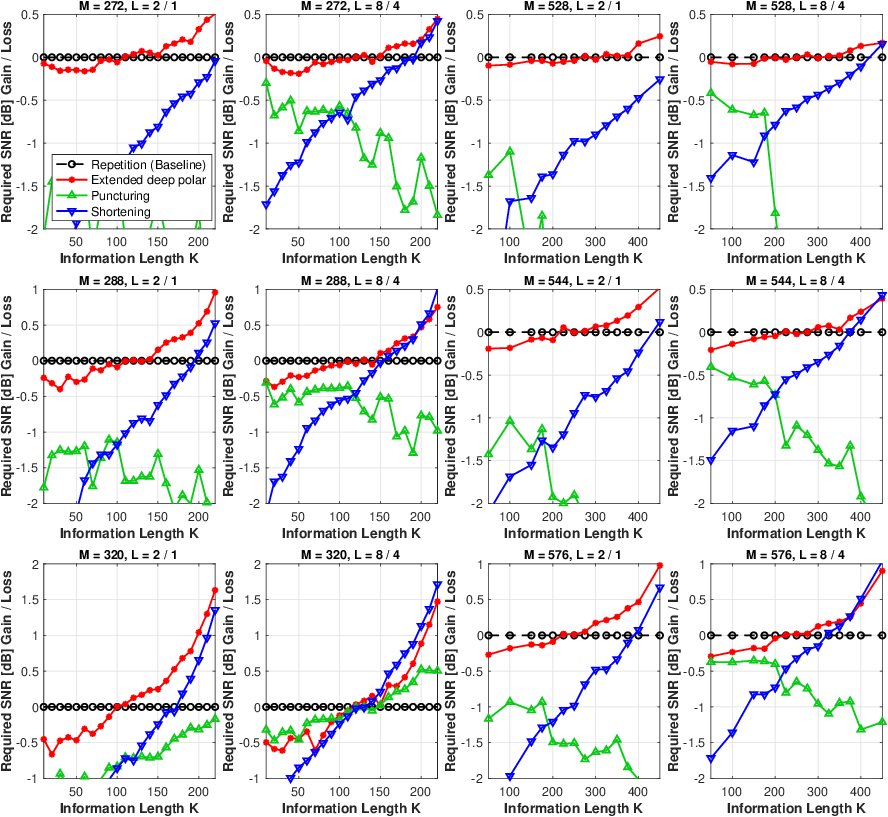}
    \caption{Required SNR [dB] to achieve BLER $10^{-3}$ for each information length $K$. A point above the zero baseline indicates a performance gain over repetition. For each $N_0 \in \{256, 512\}$, we select $1$ pre-transform size from $N_q \in \{ 16, 32, 64 \}$.}\label{fig:required-snr-single-layer}
\end{figure*}

\begin{figure*}
    \centering
    \includegraphics[width=\textwidth]{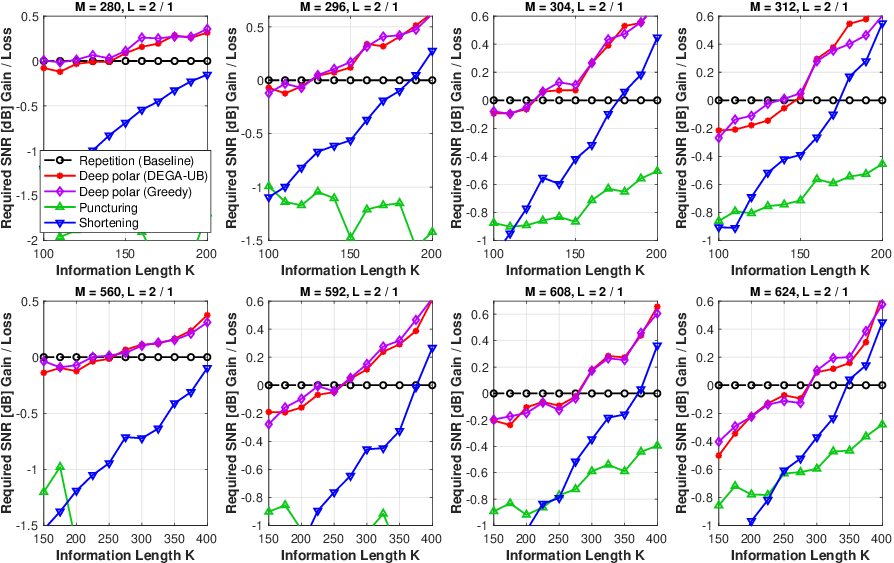}
    \caption{Required SNR [dB] to achieve BLER $10^{-3}$ for each information length $K$. A point above the zero baseline indicates a performance gain over repetition. For multi-layer configurations, at $N_0=256$, we select $2$ or $3$ pre-transform sizes from $N_q \in \{8, 16, 32\}$, and at $N_0=512$, we select $2$ or $3$ pre-transform sizes from $N_q \in \{16, 32,64\}$. For example, $280=256+16+8$ and $312=256+32+16+8$.} \label{fig:required-snr-multi-layer}
\end{figure*}

Fig.~\ref{fig:required-snr-single-layer} and \ref{fig:required-snr-multi-layer} illustrate the comparative error-correcting capabilities of four distinct rate-matched polar coding schemes: repetition, puncturing, shortening, and deep polar based extension. These figures present the minimally-required SNR to achieve the BLER of $10^{-3}$. For enhanced clarity in presentation, we establish the performance of repetition-based schemes as a baseline and quantify the relative performance gains and deficits of alternative rate-matching methods. Data points positioned above the zero baseline indicate superior performance relative to repetition-based approaches. For each information length $K$, we conduct simulations to derive BLER as a function of SNR, subsequently employing interpolation techniques to determine the corresponding SNR value at which a BLER of $10^{-3}$ is attained.

\subsubsection{Single-Layer Extended Deep Polar Codes}
 Fig.~\ref{fig:required-snr-single-layer} presents a comprehensive analysis of single-layer extended deep polar codes across diverse code rates and codeword dimensions $M \in \{272, 288, 320, 528, 544, 576\}$. For the smaller codeword configurations $M \in \{272, 288, 320\}$, we implemented a mother code size of $N_0 = 256$ for repetition and single-layer extended deep polar codes while employing a larger mother code size of $N_0 = 512$ for puncturing and shortening methods. Similarly, for larger codeword dimensions $M \in \{528, 544, 576\}$, a mother code size of $N_0 = 512$ was utilized for repetition and single-layer extended deep polar codes, contrasted with $N_0 = 1024$ for puncturing and shortening techniques.

SCL decoders with list sizes of $L=2$ and $L=8$ were deployed for extended deep polar codes and repetition-based schemes. To perform fair comparisons regarding computational complexity, memory utilization, and processing latency, reduced list sizes of $L=1$ and $L=4$ were employed for puncturing and shortening methods.

Within the figure's organizational structure, rows correspond to progressively increasing values of $M$, while columns represent variations in list size and mother code dimensions while maintaining consistent extension size $M-N_0$.

Across the spectrum of codeword lengths $M$ and list sizes examined, our deep polar-based extension consistently demonstrates superior error-correcting capabilities, particularly in scenarios characterized by modest extension sizes $M-N_0$ and constrained list sizes. Under these conditions, puncturing and shortening techniques exhibit comparatively inferior performance relative to repetition, primarily attributable to the elimination of approximately half the codeword bits. Moreover, the exceptional performance achieved with limited list sizes suggests that extended deep polar codes possess enhanced selectivity in reinforcing weaker information bits compared to alternative methods.

Each subfigure consistently demonstrates that as the code rate increases, the extended deep polar code exhibits markedly superior performance relative to conventional rate-matching techniques. These empirical findings provide compelling evidence for the superiority of our proposed method across a comprehensive range of code parameters.

\subsubsection{Multi-Layer Extended Deep Polar Codes}
Fig.~\ref{fig:required-snr-multi-layer} illustrates the performance of multi-layer extended deep polar codes across various code rates and codeword dimensions. For codeword lengths $M \in \{280, 296, 304, 312\}$, we employed a mother code size of $N_0 = 256$ for repetition and multi-layer extended deep polar codes, while utilizing $N_0 = 512$ for puncturing and shortening techniques. Similarly, for larger codeword dimensions $M \in \{560, 592, 608, 624\}$, we implemented a mother code size of $N_0 = 512$ for repetition and multi-layer extended deep polar codes, contrasted with $N_0 = 1024$ for puncturing and shortening approaches. SCL decoders with list size $L=2$ were utilized for repetition and deep polar codes, whereas a list size of $L=1$ was applied for  puncturing and shortening methods.

We investigated two alternative construction methods for extended deep polar codes: i) an exhaustive optimization procedure to identify optimal $K_q^\star$ values, and ii) a computationally efficient greedy search algorithm as detailed in Algorithm~\ref{Alg:find-K-greedy}. For the determination of $N_q$ parameters, we employed a binarization representation-based approach as previously elucidated. Within the figure's organizational framework, columns correspond to incrementally increasing values of $M$ while preserving consistent mother code dimensions, whereas rows reflect variations in the mother code size. Each row comprises three subfigures representing two-layer extended deep polar codes, complemented by a fourth subfigure illustrating three-layer extended deep polar codes.

The experimental results presented in each subfigure substantiate that our proposed extension methods retain their effectiveness in multi-layer configurations. Consistent with observations in the single-layer extension context, our proposed scheme demonstrates particularly remarkable performance when the extension magnitude $M-N_0$ is relatively modest. Additionally, the empirical evidence indicates that our greedy algorithm for designing multi-layered deep polar codes attains comparable error-correcting capabilities to exhaustive parameter searching utilizing the upper bound of SC decoding error probability obtained by DEGA, while systematically outperforming conventional rate-matching strategies predicated on repetition, puncturing, and shortening techniques.

\section{Conclusion}
We presented a novel extension method based on deep polar codes for scenarios where the desired code length is  larger than a power of two. By leveraging the hierarchical structure of deep polar codes and incorporating soft information processing through SoSCL decoding, our approach demonstrates significant performance advantages over conventional rate-matching techniques across diverse code parameters. The proposed greedy algorithm reduces design complexity from exponential to linear in the number of layers while maintaining near-optimal performance. Our method offers an efficient solution to the rate-matching problem without the complexity overhead associated with puncturing and shortening techniques, making it well-suited next-generation communication systems requiring flexible blocklenghts and high reliability.

%%%%%%
%% Appendix:
%% If needed a single appendix is created by
%%
%\appendix
%%
%% If several appendices are needed, then the command
%%
% \appendices
%%
%% in combination with further \section commands can be used.
%%%%%%

%%%%%%
%% To balance the columns at the last page of the paper use this
%% command:
%%
%\enlargethispage{-1.2cm} 
%%
%% If the balancing should occur in the middle of the references, use
%% the following trigger:
%%
% \IEEEtriggeratref{25}
%%
%% which triggers a \newpage (i.e., new column) just before the given
%% reference number. Note that you need to adapt this if you modify
%% the paper.  The "triggered" command can be changed if desired:
%%
%\IEEEtriggercmd{\enlargethispage{-20cm}}
%%
%%%%%%

% \newpage

\bibliographystyle{IEEEtran}
\bibliography{bibliography/IEEEAbbr, bibliography/coding}

\end{document}